\pdfoutput=1
\documentclass[11pt]{article} 
\usepackage{epsfig} 
\usepackage{amssymb,amsmath,amsthm,amscd}
\usepackage{latexsym} 
\usepackage{cancel}
\usepackage[dvipsnames]{xcolor}
\usepackage{amssymb}

\usepackage[margin=1.1in]{geometry}
\usepackage{float}
\usepackage{chngcntr}
\usepackage{graphicx}
\usepackage{textcomp}
\counterwithout*{equation}{section} 
 
\theoremstyle{definition}

\theoremstyle{remark} 
\newtheorem{rem}{Remark}[section]  
\def\eqref#1{(\ref{#1})} 

\newcommand {\mat}      [1] {\left[\begin{array}{#1}}
\newcommand {\rix}          {\end{array}\right]}
\newcommand {\de}      [1] {\left|\begin{array}{#1}}
\newcommand {\nt}          {\end{array}\right|}

\newcommand{\bstar}       {\begin{eqnarray*}}
\newcommand{\estar}       {\end{eqnarray*}}
\newcommand{\eqn}       {\begin{eqnarray}}
\newcommand{\enn}       {\end{eqnarray}}
\newcommand{\eq}[1]   {\begin{equation}\label{#1}}
\newcommand{\en}      {\end{equation}}

\begin{document}  
\begin{titlepage}  
\title{Effects of diffusion coefficients on reversal potentials in ionic channels}  
\author{Bob Eisenberg \footnote{Department of Molecular Biophysics and Physiology, Rush Medical Center, 1759 Harrison St.,
Chicago, IL 60612 (beisenbe@rush.edu).}\;, Weishi Liu\footnote{Department of Mathematics, University of Kansas, 
1460 Jayhawk Blvd., Room 405,
Lawrence, Kansas 66045, USA ({\tt wsliu@ku.edu}).}\;, and  Hamid Mofidi\footnote{Department of Mathematics, University of Kansas, 
1460 Jayhawk Blvd., Room 405,
Lawrence, Kansas 66045, USA ({\tt h.mofidi@ku.edu}).  
}}
\date{\today} 

\end{titlepage}

\maketitle    

\begin{abstract} 
{
In this work, the dependence of reversal potentials and zero-current fluxes on diffusion coefficients are examined for ionic flows through membrane channels. The study is conducted for the setup of a simple structure defined by the profile of permanent charges with two mobile ion species, one positively charged (cation) and one negatively charged (anion). Numerical observations are obtained from analytical results established using geometric singular perturbation analysis of classical Poisson-Nernst-Planck models. For 1:1 ionic mixtures with arbitrary diffusion coefficients, Mofidi and Liu [{\em arXiv:1909.01192 }]  conducted a rigorous mathematical analysis and derived an equation for reversal potentials that, in its particular case, can be compared to Goldman-Hodgkin-Katz equation. We summarize and extend these results with numerical observations for biological relevant situations. The numerical investigations on profiles of the electrochemical potentials, ion concentrations, and electrical potential across ion channels are also presented for the zero-current case. Moreover,
the behavior of current and fluxes with respect to voltages and permanent charges  are investigated.  
In the opinion of the authors, many results in the paper are not intuitive, and it is difficult, if not impossible, to see all cases without investigations of this type.
}

 \end{abstract} 


\section{Introduction.} 
{
Ion channels are proteins  found in cell membranes that create tiny openings in the membrane to allow cells to communicate with each other and with the outside to transform signals and to conduct tasks together \cite{BNVHEG, Eis00}.
They have an aqueous pore, that becomes accessible to ions after a change in the protein structure that makes ion channels  open.
Ion channels permit the selective passage of charged ions formed from dissolved salts, including sodium, potassium, calcium, and chloride ions that carry electrical current in and out of the cell.
}


To unravel how ion channels operate, one needs to understand the physical structure of ion channels, which is defined by the channel shape and the permanent charge.
The shape of a typical ion channel is often approximated as a cylindrical-like domain.
Within a large class of ion channels, amino acid side chains are distributed mainly over a ``short" and ``narrow" portion of the channel, with acidic side chains contributing permanent negative charges and basic side chains contributing permanent positive charges. The spatial distribution of side chains in a specific channel defines the permanent charge of the channel.

The most basic function of ion channels is to regulate the permeability of membranes for a given
species of ions and to select the types of ions and to facilitate and modulate the diffusion of ions across cell membranes. 
At present, these permeation and selectivity properties of  ion channels  are usually determined from the current-voltage (I-V) relations measured experimentally (\cite{Eis00, Gillespie08}). 
  Individual fluxes carry more information than the current, but it is expensive and challenging to measure them (\cite{HK55, JEL19}). 
The I-V relation defines the function of the channel structure, namely the ionic transport through the channel. That transport depends on driving forces expressed mathematically as boundary conditions. 
The multi-scale feature of the problem with multiple physical parameters allows the system to have great flexibility and to exhibit vibrant phenomena/behaviors -- a great advantage of ``natural devices" (\cite{Eis}). On the other hand, the same multi-scale feature with  multiple physical parameters  presents an extremely challenging task for anyone to extract meaningful information from experimental data, also given the fact that the internal dynamics cannot be measured with present techniques. The general inverse problem is challenging, although specific inverse problems have been successfully solved with surprisingly little difficulty using standard methods and software packages (\cite{BEE}).

 To understand the importance of the relation of current and permanent charges, i.e., I-Q, we point out that the role of permanent charges in ionic channels is similar to the role of  doping profiles in semiconductor devices. Semiconductor devices are similar to ionic channels in the way that they both use atomic-scale structures to control macroscopic flows from one reservoir to another. Ions move much as quasi-particles move in semiconductors.
Semiconductor technology controls the migration and diffusion of quasi-particles of charge in transistors and integrated circuits. Doping is the process of adding impurities into intrinsic semiconductors to modulate its electrical, optical, and structural properties (\cite{Rous90, Warn01}).


{
Ion channels are almost always passive and do not require a source of chemical energy (e.g., ATP hydrolysis)  in order to operate. Rather they allow ions to flow passively driven by a combination of the transmembrane electrical potential and the ion concentration gradient across the membrane.  For fixed other physical quantities, the total current $\mathcal{I}=\mathcal{I}(\mathcal{V}, \mathcal{Q})$  depends on the transmembrane potential $\mathcal{V}$ and the permanent charge $\mathcal{Q}$. For fixed $\mathcal{Q}$, {\em a reversal potential} $\mathcal{V}=\mathcal{V}_{rev}(\mathcal{Q})$ is a transmembrane potential that produces zero current $ \mathcal{I}(\mathcal{V}_{rev}(\mathcal{Q}),\mathcal{Q})=0$. Similarly, for fixed transmembrane potential $\mathcal{V}$, {\em a reversal permanent charge} $\mathcal{Q}=\mathcal{Q}_{rev}(\mathcal{V})$  is a permanent charge that produces zero current $ \mathcal{I}(\mathcal{V}, \mathcal{Q}_{rev}(\mathcal{V}))=0$.
}


The {\em Goldman-Hodgkin-Katz (GHK) equation} for reversal potentials involving multiple ion species (\cite{Goldman, HK49}) is used to determine the reversal potential across ion channels. The GHK equation is an extension of the Nernst equation -- the latter is for one ion species.
The classical derivations were based on the incorrect assumption that the electric potential $\Phi(X)$ is linear in $X$ -- the coordinate along the longitude of the channel. This assumption is particularly unfortunate because it is the change in the shape of the electrical potential that is responsible for so much of the fascinating behaviors   transistors or ionic systems (\cite{Eis96, Eis96(2), MRS90, Sel84, Sho50, VGK10}).
 There was no substitute for their equations until  authors of \cite{ELX, ML19} recently offered  equations derived from self-consistent PNP systems, to be the best of our knowledge.

There have been some achievements recently in analyzing the PNP models for ionic flows through ion channels (\cite{EL,JEL19, JL12, JLZ15, Liu05, Liu09}, etc.). Although mathematical analysis plays a powerful and unique role to explain mechanisms of observed biological phenomena and to discover new phenomena, numerical results are needed to fit actual experimental data and study cases where analytical solutions do not exist.  Furthermore,  numerical observations give clues for more theoretical investigations. Indeed, numerical and analytical studies are linked; any progress in one catalyzes work in the other.

The numerical results, throughout the paper, are gained from the algebraic systems \eqref{G1G2Sys},  \eqref{VonQf}, \eqref{N-GHK} and  \eqref{IJ}, obtained from reduced matching systems  of analytical results in \cite{ML19} and \cite{EL}.  The nonlinear algebraic systems are then solved by the \textsc{Matlab}\textsuperscript{\textregistered} function {\em fsolve} that uses the trust-region dogleg algorithm. The trust-region algorithm is a subspace trust-region method and is based on the interior-reflective Newton method described in \cite{CL96}.
Our numerical results indicate that current-voltage and current-permanent charge and even zero-current relations depend on a rich interplay of boundary conditions and the channel geometry arising from the mathematical properties analyzed in  (\cite{EL, JLZ15, ML19, ZEL}).

This paper is organized as follows. The classical PNP model is provided for ionic flows in Section \ref{sec-PNP} to prepare the stage for investigations on the next sections. 
In Section \ref{sec-ZeroCur}, we  study zero current problems to investigate zero-current fluxes, and reversal potentials $\mathcal{V}_{rev}$.  In particular, we compare a special case of the reversal potential with the GHK equation. Some other numerical observations are also provided to study profiles of relevant physical quantities in Section \ref{sec-PhysQuan}.
 In Section \ref{sec-CurVol}, we first recall the analytical results in \cite{EL} when diffusion constants are also involved. Then numerical observations are provided to examine behaviors of current, voltage, and permanent charge with respect to each other in some general cases. 
 Some concluding remarks are provided in Section \ref{sec-Conc}.


\subsection{Poisson-Nernst-Planck models for ionic flows.}\label{sec-PNP}
\setcounter{equation}{0}
The PNP system of equations has been analyzed mathematically to some extent, but the equations have been simulated and computed to a much larger extent (\cite{BCE, CE, ChK, HCE, IR}). One can see from these simulations that macroscopic reservoirs must be included in the mathematical formulation to describe the actual behavior of channels (\cite{GNE, NCE}).
For an ionic mixture of $n$ ion species, PNP type model is, for $k=1,2,..., n$,
\begin{align}\label{PNP}
\begin{split} 
 \text{Poisson:} \quad & \nabla \cdot\Big( \varepsilon_r(X) \varepsilon_0 \nabla \Phi \Big) = -e_0 \Big( \sum_{s=1}^n z_s C_s + \mathcal{Q}(X)\Big), \\
 \text{Nernst-Planck:} \quad & \partial_t C_k + \nabla \cdot \mathcal{J}_k =0, \quad - \mathcal{J}_k =\dfrac{1}{k_BT} \mathcal{D}_k(X) C_k \nabla \mu_k,
 \end{split}
\end{align} 
where $X \in \Omega$ with $\Omega$ being a three-dimensional cylindrical-like domain representing the channel of length $\hat{L} ~(nm = \hat{L} \times 10^{-9}m)$, ${\cal Q}(X)$ is the permanent charge density of the channel (with unit $1M=1\frac{mol}{L}=10^3\frac{mol}{m^3}$ ), $\varepsilon_r(X)$ is the  relative dielectric coefficient (with unit 1), $\varepsilon_0 ( \approx 8.854187817620 \times 10^{-12}~ F m^{-1})$ is the vacuum permittivity, $e_0$($ \approx 1.60217646 \times 10^{-19}$ $C=$coulomb) is the elementary charge, $k_B(=1.380648813 \times 10^{-23} JK^{-1})$  is the Boltzmann constant, $T$ is the absolute temperature ($T\approx 273.16~ K=$kelvin, for water);  $\Phi$ is the electric potential (with the unit $V=\text{Volt}=JC^{-1}$), and, for  the $k$-th ion species, $C_k$ is the concentration (with unit   $M$), $z_k$ is  the valence (the number of charges per particle with unit $1$), $\mu_k$ is the electrochemical potential depending on $\Phi$ and $C_k$ (with unit $J=CV$). The flux density ${\cal J}_k(X)$ (with unit $Hz =\frac{1}{s}$) is the number of particles across each cross-section in per unit time, ${\cal D}_k(X)$ is the diffusion coefficient (with unit $m^2/s$), and $n$ is the number of distinct types of ion species (with unit $1$).  

Ion channels have narrow cross-sections relative to their lengths. Therefore, three-dimensional PNP type models can be reduced to quasi-one-dimensional models.  The authors of \cite{NE} first offered a reduced form, and for a particular case, the reduction is precisely verified in \cite{LW10}. The quasi-one-dimensional steady-state PNP type is, for $k = 1, 2, ..., n,$
\begin{align}\label{1dPNP}
\begin{split} 
 \frac{1}{ A(X)}  \frac{d}{dX}\left({\varepsilon}_r(X)\varepsilon_0 A(X) \frac{d \Phi}{d X}\right)=&-e_0\left( \sum_{s=1}^nz_s C_s + \mathcal{Q}(X)\right), \\ 
\frac{d \mathcal{J}_k}{d X}  =0, \quad  -\mathcal{J}_k=& \dfrac{1}{k_BT}\mathcal{D}_k(X)A(X)C_k\frac{d \mu_k}{dX},   
\end{split} 
\end{align} 
where $A(X)$ is the area of cross-section of the channel over location $X$.
Equipped with system (\ref{1dPNP}),  we impose the following   boundary conditions, for $k=1,2,\cdots, n$, 
\begin{equation}\label{BV} 
\Phi(0)={\cal V},  \quad C_k(0)=L_k>0; \quad \Phi(\hat{L})=0,\quad C_k(\hat{L})=R_k>0.
\end{equation} 

\noindent One often uses the electroneutrality conditions on the boundary concentrations because the solutions are made from electroneutral solid salts,
 \begin{align}\label{neutral}
 \sum_{s=1}^nz_sL_s= \sum_{s=1}^nz_sR_s=0.
 \end{align}
The electrochemical potential ${\mu}_k(X)$ for the $k$-th ion species consists of the ideal component ${\mu}_k^{id}(X)$ and the excess component ${\mu}_k^{ex}(X)$, i.e., ${\mu}_k(X)={\mu}_k^{id}(X)+{\mu}_k^{ex}(X)$. The  excess electrochemical potential ${\mu}_k^{ex}(X)$   accounts for the finite size effect of ions. 
It is needed whenever concentrations exceed say $50 mM$, as they almost always do in technological and biological situations. The   classical  PNP model    only deals with  the ideal component ${\mu}_k^{id}(X)$, which   reflects the collision between ion particles and  water molecules and ignores the size of   ions; that is,
\begin{equation}\label{Ideal}
{\mu}_k(X)={\mu}_k^{id}(X)= z_k e_0 \Phi(X)+k_B T \ln \frac{C_k(X)}{C_0},
\end{equation}
where $C_0$ is a characteristic concentration of the problems, and one may consider,
\begin{equation}\label{CharConc}
C_0=\max_{1\le k\le n}\big\{ L_k, R_k, \sup_{X\in[0,\hat{L}]}|{\cal Q}(X)|\big\}.
\end{equation}


For given $\mathcal{V}$, $\mathcal{Q}(X)$, $L_k$'s and $R_k$'s,   if  $(\Phi(X), C_k(X), \mathcal{J}_k)$ is a solution of the boundary value problem (BVP) of (\ref{1dPNP}) and (\ref{BV}), then  the electric current $\mathcal{I} $ is   
 \begin{align}\label{IV}
\mathcal{I}= e_0 \sum_{s=1}^nz_s \mathcal{J}_s.
 \end{align}

For an analysis of the boundary value problem (\ref{1dPNP}) and (\ref{BV}), we work on a dimensionless form. 
Set 
\[{\cal D}_0=\max_{1\le k\le n}\{\sup_{X\in [0,\hat{L}]}{\cal D}_k(X)\}\;\mbox{ and }\; \bar{\varepsilon}_r=\sup_{X\in [0,\hat{L}]} \varepsilon_r(X).\]
Let
\begin{equation}\label{DimToDimless}
\begin{aligned}
&\varepsilon^2=\frac{\bar{\varepsilon}_r\varepsilon_0k_BT}{e_0^2\hat{L}^2C_0},\quad
  \hat{\varepsilon}_r(x)=\frac{\varepsilon_r(X)}{\bar{\varepsilon}_r},\quad x=\frac{X}{\hat{L}},\quad  h(x)=\frac{A(X)}{\hat{L}^2},\quad D_k(x)=\frac{{\cal D}_k(X)}{{\cal D}_0},\\
  & Q(x)=\frac{{\cal Q}(X)}{C_0},  \quad \phi(x)=\frac{e_0}{k_BT}\Phi(X), \quad c_k(x)=\frac{C_k(X)}{C_0},\quad \hat{\mu}_k=\frac{1}{k_BT}\mu_k, \quad J_k=\frac{{\cal J}_k}{ \hat{L} C_0{\cal D}_0}. 
 \end{aligned}
\end{equation}
In terms of the new variables, the  BVP of  (\ref{1dPNP}) and (\ref{BV}) become, for $k=1,2,\cdots,n$,
\begin{align}\label{1dPNPdim}
\begin{split} 
 \frac{\varepsilon^2}{ h(x)}  \frac{d}{dx}\left(\hat{\varepsilon}_r(x)h (x)\frac{d}{dx}\phi\right)=&- \sum_{s=1}^nz_s c_s -  Q(x), \\ 
\frac{d J_k}{dx}  =0, \quad  -J_k=&h (x)D_k(x)c_k\frac{d  }{dx}\hat{\mu}_k ,   
\end{split} 
\end{align} 
with the boundary conditions  
\begin{equation}\label{1dBV} 
\phi(0)=V=\frac{e_0}{k_BT}{\cal V},  \quad c_k(0)=l_k=\frac{L_k}{C_0}; \quad \phi(1)=0,\quad c_k(1)=r_k=\frac{R_k}{C_0}.
\end{equation} 
\begin{rem}\label{rem-dim2}
{\em The actual dimensional  forms of quantities have been used for all figures throughout the paper, i.e., for $k=1,2,$
\begin{equation}
\begin{aligned}
C_k =& C_0 c_k~ (M=Molar),\quad \quad \mathcal{Q}= C_0 Q~ (M), \\
\Phi=& \frac{k_B T}{e_0}\phi\times 10^3 ~(mV), \quad 
 ~\mathcal{V} = \frac{k_BT}{e_0} V~ (mV),\\
 \mu_k=& \big( e_0 \Phi+ k_B  T \ln(C_k/C_0) \big) \times 10^{21}~  (zJ)
 = \phi(x) + \ln c_k \quad (in~~ k_BT),\\
 \mathcal{J}_k =& \hat{L} C_0 \mathcal{D}_0 J_k \times 10^{-6} ~(MHz),\quad
 \mathcal{I} = \hat{L} C_0\mathcal{D}_0 e_0 I \times 10^{12} ~ (pA),
\end{aligned}
\end{equation}
where we take
$C_0 = 10 M,  \hat{L} = 2.5 nm,$ and $
\mathcal{D}_0 =2.032\times 10^{-9} ~ m^2/s,
$
 and, for diffusion constants (\cite{LE14}), for $k=1,2,$ 
\begin{equation}\label{Dif-Coef}
\begin{aligned}
\mathcal{D}_k =& 1.334 \times 10^{-9} ~ m^2/s~ \text{for Na}^+,  \; \text{or}\\
\mathcal{D}_k =& 2.032 \times 10^{-9} ~ m^2/s~ \text{for Cl}^-, \; ~\text{or}\\
\mathcal{D}_k =& 0.792 \times 10^{-9} ~m^2/s~ \text{for Ca}^{2+}.
\end{aligned}
\end{equation}
}
\end{rem}

\subsection{Setup of the problem.}
We give the precise setup of the problem considered in this work. More precisely, we assume
{\em \begin{itemize}
 \item[(A0)]  The ionic mixture consists of two ion species  with valences $z_1=-z_2=1$;
  \item[(A1)] $D_k(x)=D_k$  for $k=1,2$ is a constant and  $\hat{\varepsilon}(x)=1$;
\item[(A2)] Electroneutrality boundary conditions (\ref{neutral}) hold;
\item[(A3)] The permanent charge $Q$ is piecewise constant  with one nonzero region; that is, for a partition $0<a<b<1$ of   $[0,1]$, 
 \begin{align}\label{Q}
 Q(x)=\left\{\begin{array}{ll}
 Q_1=Q_3=0, & x\in (0,a)\cup (b,1),\\
 Q_2, & x\in (a,b),
 \end{array}\right.
 \end{align}   
 where $Q_2$ is a constant.
 \end{itemize}}
{  We assume that $\varepsilon>0$ in system \eqref{1dPNPdim} is small.  The assumption is reasonable since, if   $ \hat{L}= 2.5nm(=2.5 \times 10^{-9}m)$ and $C_0 = 10M$, then $\varepsilon \approx 10^{-3}$ (\cite{EL17}). 
The assumption that $\varepsilon$ is small enables one to treat the BVP \eqref{1dPNPdim} and \eqref{1dBV} of the dimensionless problem as a singularly perturbed problem. 

A geometric singular perturbation framework for analyzing BVP of cPNP systems was developed first in \cite{EL, Liu05} for ionic mixtures with two types of ion species. This general framework was extended to arbitrary number of types of ion species  successfully  only when two special structures of the PNP system were revealed (\cite{Liu09}). One special structure is {\em a complete set of integrals} for the limit fast (or inner) system that allows a detailed analysis of singular layer component of the full problem. The other is {\em a state-dependent scaling} of the independent variable that turns the nonlinear limit slow (or outer) system to a linear system with constant coefficients. The coefficients do depend on unknown flux densities to be determined as a part of the whole problem, and this is the mathematical reason for the rich dynamics of the problem. As a consequence of the framework, the existence, multiplicity, and spatial profiles of the singular orbits are reduced to a system of nonlinear algebraic equations that involves all relevant quantities altogether -- the physical evidence why this framework works since ``Everything interacts with everything else" in determining ion channel functions. This geometric framework with its extensions to include ion size effects to some extents in \cite{JL12, LTZ12,  SL18} has produced a number of results that are central to ion channel properties (\cite{JEL19,JLZ15, LLYZ13, Liu18,ML19,ZEL}); for example, it was shown in \cite{JLZ15} that, in order to optimize effects of the permanent charge, the channel should have a short and narrow neck within which the permanent charge is confined; and, it was shown in \cite{ZEL} that, large permanent charge is responsible for the declining phenomenon -- decreasing flux with increasing transmembrane electrochemical potential.
  We refer the readers to the aforementioned papers for more details on geometric singular perturbation framework for PNP as well as concrete applications to ion channel problems.

In this paper,  we will apply some results and follow the notations in \cite{EL} and \cite{ML19} for analytical results where the quantities are all in their dimensionless forms. Besides, for simplicity, we use the letters $l$, $r$ and $Q_0$ where
$ l_1=l_2=l$, $r_1=r_2=r$, $Q=2Q_0$.  Also, for definiteness, we choose
$a = {1}/{3}$, $b ={2}/{3}$ and $h(x)=1$. 
}

\begin{rem}\label{rem-dim1}
{\em We remind the readers that the quantities $V,l,r, c_k,Q, \phi, \hat{\mu}_k, J_k, D_k$ and $I$ are dimensionless quantities corresponding to the dimensional quantities  $\mathcal{V},L,R, C_k,\mathcal{Q}, \Phi, \mu_k, \mathcal{J}_k, \mathcal{D}_k$ and $\mathcal{I}$, respectively, obtained from \eqref{DimToDimless}. We switch from dimensional form to the dimensionless form and vice versa several times throughout the paper.  }
\end{rem}


\section{Zero current problems with general diffusion constants.}\label{sec-ZeroCur}
\setcounter{equation}{0}
In this section, we study when and how fluxes and membrane potential produce current reversal. Throughout this section, in order to express the effects of diffusion constants on zero-current flux and reversal potential, we study and compare the results for different cases of diffusion constants where  $\mathcal{D}_1=\mathcal{D}_2$ and where $\mathcal{D}_1 \neq \mathcal{D}_2$,  to indicate and emphasize the differences.


Diffusion is the phenomenon through which solute particles spread or mix as a result of their potential energy. It is a spontaneous process that acts to eliminate differences in concentration and eventually leads a given mixture to a state of uniform composition. 
Fick's first law \cite{Fick}  describes diffusion of uncharged particles by $\frac{\partial c}{\partial t}= \mathcal{D}\frac{\partial^2 c}{\partial x^2}$, where $c$ is the concentration, $\mathcal{D}$ is the diffusion constant and $t$ is time.
 Frequently, the determination of diffusion constants involves measuring sets of simultaneous values of $t$, $c$, and  $x$. These measured values are then applied to a solution of Fick's law to get the diffusion constants. 
Many techniques are available for the determination of diffusion constants of ions (charge particles) in aqueous solutions (\cite{BF80,GA82,BHW66,LE14,Smith69}, etc.). 
When diffusion constants are equal, classical electrochemistry tells that many electrical phenomena ``disappear" altogether, e.g., the ``liquid junction" is zero. If the diffusion constants of potassium and chloride are equal, classical electrochemistry says that KCl acts nearly as an uncharged species. Indeed, that is the basis for the saturated KCl salt bridge used in a broad range of electrochemical experiments for many years. Therefore, the equal diffusion constants case is quite degenerate.

Experimental measurements are exclusively performed under isothermal conditions to avoid deviation of $\mathcal{D}$ values. Nevertheless, even diffusion constants of certain ionic species may differ from one method to another, even when all other parameters are held constant.  Many   things becomes much more complicated mathematically when the diffusion constants are not equal, however. This complexity is what makes many biological and technological devices valuable. Some kinds of selectivity depend on the non-equality of diffusion constants, as well.

Applying GSP theory to cPNP equations  \eqref{PNP}  for two ion species with diffusion constants $\mathcal{D}_k, ~k=1,2$,  the authors of \cite{ML19} obtained an algebraic matching system with eleven equations and eleven unknowns for zero current problems and singular orbits on $[0,1]$. In order to be able to analyze it, they reduced the matching system to the case where two ion valences satisfy $z_1=-z_2$. 
It follows from \cite{ML19} that the reduced matching system for zero current ${I}= J_1 -J_2 =0$   when $z_1=-z_2=1$ is,
\begin{align}\label{G1G2Sys}
G_1(A, Q_0,\theta )=V\;\mbox{ and }\; G_2(A,Q_0,\theta )=0,
\end{align}
where
\begin{equation}\label{G}
\begin{aligned}
G_1(A, Q_0,\theta )=& \theta  \Big(\ln\dfrac{S_a + \theta Q_0}{S_b + \theta  Q_0} + \ln\dfrac{l}{r}\Big)  - (1+\theta )\ln \dfrac{A}{B} + \ln \dfrac{S_a -Q_0}{S_b -Q_0},\\
G_2(A, Q_0,\theta )=& \theta Q_0\ln\dfrac{S_a+\theta Q_0}{S_b+\theta Q_0}-N.
\end{aligned}
\end{equation}	 
In above, $\theta =\frac{D_2- D_1}{D_2 + D_1}$, and $A$   is the geometric mean of concentrations at $x=a$, that is,  
\begin{equation}\label{A}
A= \sqrt{c_1(a)c_2(a)}.
\end{equation}
Moreover,  
\begin{equation}\label{Nab}
  S_a= \sqrt{Q_0^2+A^2}, \quad S_b= \sqrt{Q_0^2+B^2}, 
 \quad N = A-l  + S_a -S_b.
\end{equation}

\noindent In what follows, the numerical observations help us have a better understanding of the zero-current problems and complete the analytical results obtained in \cite{ML19}.

\subsection{Zero-current flux.}
We aim to clarify the relationships of ion fluxes with permanent charge and diffusion constants when current is zero. Recall that fluxes $J_1$ and $J_2$ are equal for this case and let $J$ denotes it.  
For any permanent charge $Q=2Q_0$, once a solution $(A,V)$ of (\ref{G1G2Sys})   is obtained, it follows from matching equations (see Appendix in \cite{ML19})  that  $ J$ is given by 
\begin{equation}\label{JJ}
J  = -\dfrac{6D_1 D_2(A- l)}{(D_1 + D_2)} = -\dfrac{6D_1 D_2(r- B)}{(D_1 + D_2) }.
\end{equation}
\medskip

{
\noindent \underline{\textbf{Sign of zero-current flux $\mathcal{J}$.}}
  It was observed in \cite{ELX} that the Nernst-Planck equation in \eqref{1dPNPdim} (with dimensionless quantities) gives, for $k=1,2$,
\begin{align}\label{JonP}
\dfrac{J_k}{D_k} \int_0^1\frac{1}{h(x)c_k(x)}dx=z_kV+\ln \frac{l}{r}.
\end{align}
Therefore, the sign of flux $J_k$ depends only on the boundary conditions $l$, $r$ and $V$.
Note that \eqref{JonP} holds for any condition, not just zero-current condition.  

 For zero-current problem, $V=V_{rev}$ depends on $l$, $r$, $D_1$, $D_2$, and $Q$ as well in general. So the sign of zero-current flux $J$ seems to depend on all quantities and to be difficult to figure out.
It is not the case. A  consequence of \eqref{JJ} together with the results in Theorem $3.4$ in \cite{ML19} is that:  
\medskip

{\em The zero-current flux $J$ has the same sign as that of $l-r$}. 

\medskip

The latter follows directly from  a statement in Theorem $3.4$ in \cite{ML19}  that, for zero-current, $l-A$ has the same sign as that of $l-r$.
  That is consistent with observations in Figure \ref{fig-JDvari} where $\mathcal{D}_1= 1.334 \times 10^{-9} ~ m^2/s \text{for Na}^+$ is fixed, and $\mathcal{D}_2$ varies from the same value (i.e., $\rho =1$), to $\mathcal{D}_2= 2.032 \times 10^{-9} ~ m^2/s~ \text{for Cl}^-$, (i.e., $\rho >1$), to a random large value (i.e., $\rho \gg 1$).

\begin{figure}[H]\label{fig-JDvari}
\centerline{\epsfxsize=3.0in
\epsfbox{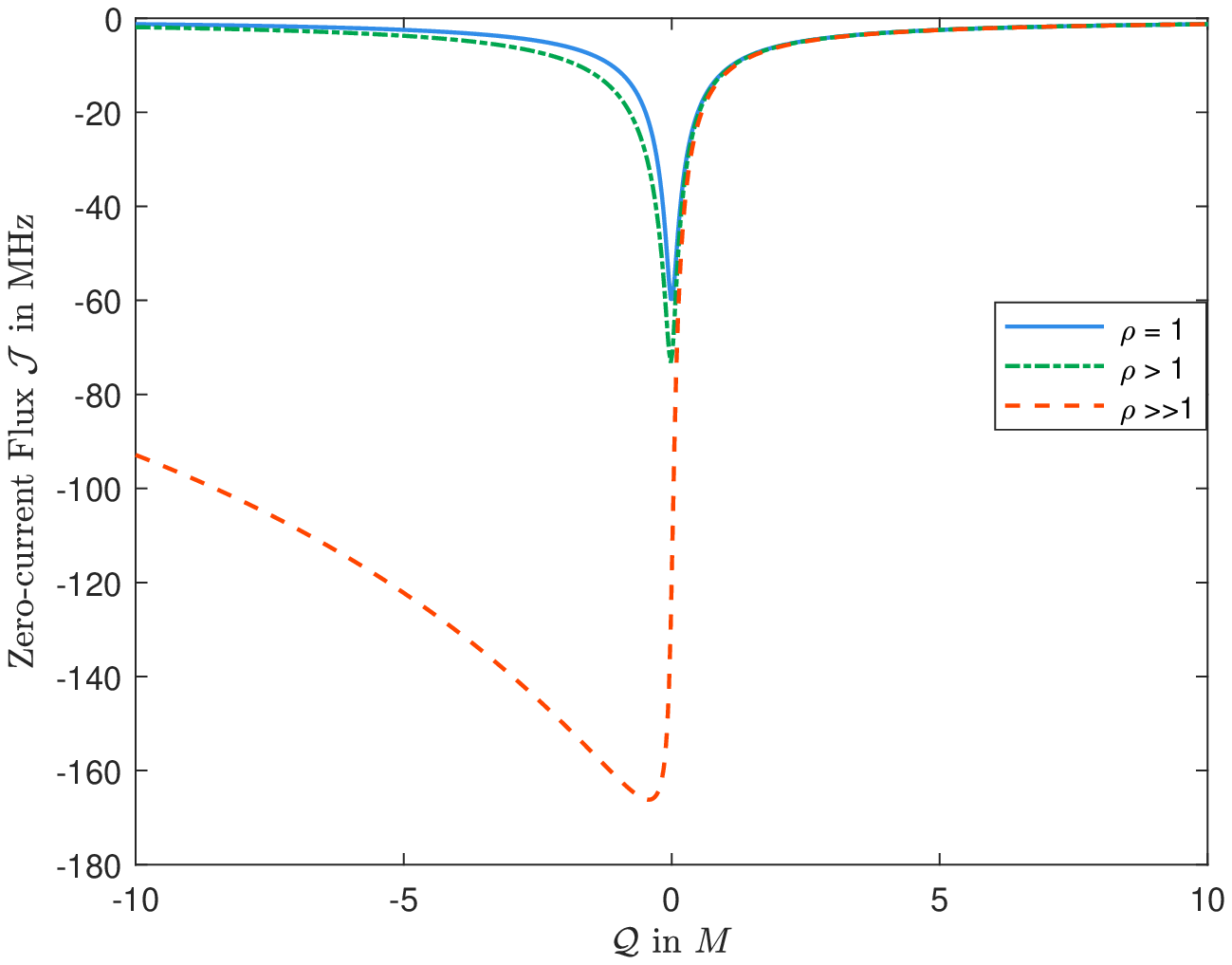} 
\epsfxsize=3.0in
\epsfbox{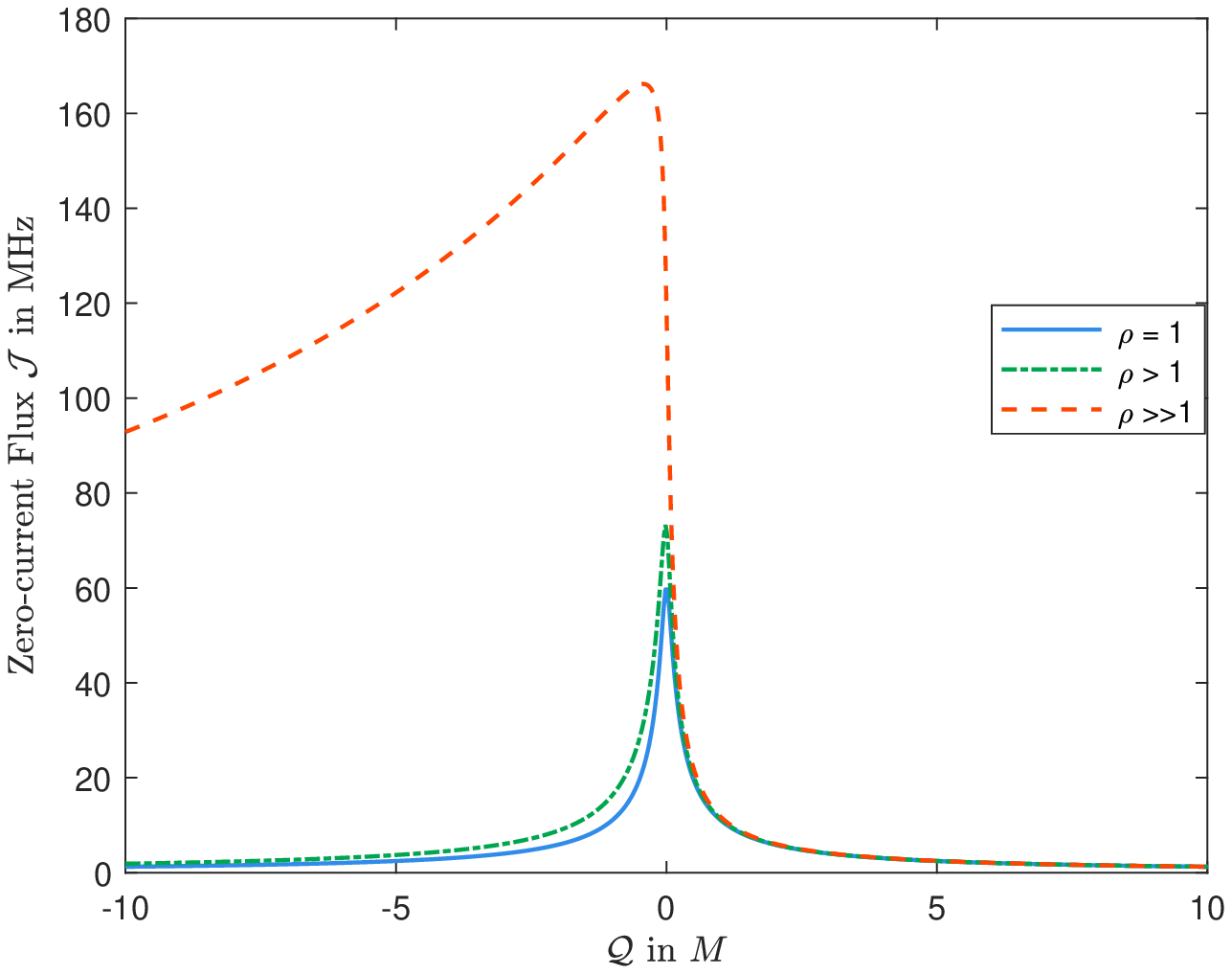} 
  }
\caption{\em The function $\mathcal{J}=\mathcal{J}(\mathcal{Q})$  for various values of $\rho = D_2 / D_1$: The left panel for  $L=2 mM$ and  $R=5 mM$; the right panel for $L= 5 mM$ and $ R= 2 mM$.}
\label{fig-JDvari}
\end{figure}


\noindent \underline{\textbf{Dependence of zero-current flux $\mathcal{J}$ on $\mathcal{Q}$ and $\mathcal{D}_k$'s.}}
Concerning the dependence of the zero-current flux $\mathcal{J}$ on $\mathcal{Q}$, we have the following.

 (i) {\em If $\mathcal{D}_1=\mathcal{D}_2$, then the zero-current flux $\mathcal{J}$ is an even function in $\mathcal{Q}$, and it is monotonically increasing for $\mathcal{Q}>0$.} 
 
\noindent In this case, $\theta=0$ and hence, it follows from \eqref{G1G2Sys}--\eqref{A} that $A$ can be determined from $N=0$. Since $N$ is an even function in $Q_0$,  $A$ is an even function in $Q_0$, and so is the zero-current flux $\mathcal{J}$ from \eqref{JJ}. 

 (ii) {\em If $\mathcal{D}_1\neq \mathcal{D}_2$,  then the zero-current flux $\mathcal{J}$ is not an even function in $\mathcal{Q}$ and the monotonicity of the zero-current flux $\mathcal{J}$  in $\mathcal{Q}$ seems to be more complicated.}
 
\noindent In this case,  it can be seen that  $G_2$ in (\ref{G}) is not an even function in $\mathcal{Q}$, and hence, the zero-current flux $\mathcal{J}$ is not.    
We would like to point out that, it follows from \cite{ZEL},  for fixed $\rho = \mathcal{D}_2/\mathcal{D}_1$, no matter how large, one always has $\mathcal{J}\to 0$ as $\mathcal{Q}\to \pm \infty$, that is  consistent with the observations in Figure \ref{fig-JDvari}.


(iii)  Another fascinating result is that {\em  for small values of $\mathcal{Q}$, the magnitude of $\rho = \mathcal{D}_2/ \mathcal{D}_1$ affects the  monotonicity of the zero-current flux $J$.}

\noindent In this case, if one fixes $\mathcal{D}_1$, and let $\mathcal{D}_2$ increases from small values to $\mathcal{D}_2 \to \infty$, (i.e.,  $\rho \to \infty$), then it follows from \eqref{JJ} that there is a meaningful change in the monotonicity of flux, for small values of $\mathcal{Q}$, that is not intuitive.
Let us consider the case where $L<R$ and $\mathcal{Q}<0$ is small. It follows from \eqref{G1G2Sys} and \eqref{G} that when  $\mathcal{Q}$ increases, for the geometric mean of concentrations $A$, one obtains

(a)~~$A$ increases if $\rho \approx 1$ (that is $\theta \approx 0$), and consequently the zero-current flux $\mathcal{J}$ decreases;

(b)~$A$ decreases if $\rho \gg 1$ (that is $\theta \gg 1$), and hence, the zero-current flux $\mathcal{J}$ increases.

  Thus, depending on the size of $\rho$, the zero-current flux $\mathcal{J}$ may increase or decrease for $\mathcal{Q}<0$ small, which is also consistent with the observations in Figure \ref{fig-JDvari}. The analysis for the   case with $L>R$ is similar.


\subsection{Reversal potential $V_{rev}$.}\label{revP}
Experimentalists have long identified reversal potential as an essential characteristic of ion channels \cite{Hod64, Hox63}.
Reversal potential is the potential at which the current reverses direction, i.e., $\mathcal{V} = \Phi(0)-\Phi(\hat{L})$ that produces zero current $\mathcal{I}$. 
Using dimensionless form of quantities (see Remark \ref{rem-dim1}), it follows from \eqref{G1G2Sys} and \eqref{G}, (where there are two ion species with valences $z_1=-z_2=1$) that  for  general permanent charge $Q=2Q_0\neq 0$ with arbitrary diffusion constants (\cite{ML19}), the reversal potential is,
\begin{align}\label{VonQf}
{V}_{rev} = \theta   \big(\ln\dfrac{S_a + \theta Q_0}{S_b + \theta  Q_0} + \ln\dfrac{l}{r}\big)  - (1+\theta ) \ln \dfrac{A(Q_0,\theta )}{B(Q_0,\theta )} + \ln \dfrac{S_a -Q_0}{S_b -Q_0}.
\end{align}


{
\noindent \underline{\textbf{Range of Reversal potential $V_{rev}$.}} 
For fixed $l$, $r$, and for any  given $Q= 2Q_0$, it follows from Theorem 4.2 in \cite{ML19} that  there exists a unique reversal potential ${V}_{rev}$ such that  ${V}_{rev}  \leq |\ln \frac{l}{r} |$. As   $Q \to \pm \infty$, then ${V}_{rev}$ gets close to the  boundary values, i.e. $ {V}_{rev} \to \pm \ln \frac{l}{r}$. 
}
\medskip


\noindent \underline{\textbf{Zero reversal potential.}} One particular case is when the reversal potential is zero.
To examine under what conditions one obtains ${V}_{rev}=0$, it follows from Theorem 4.2. in \cite{ML19} that,

(i)~ {\em if ${D}_1 = {D}_2$, then ${V}_{rev}({Q})=0$ for ${Q}=0$},

(ii) {\em if ${D}_1<{D}_2$, then there is a  ${Q}<0$, such that  ${V}_{rev}({Q})=0$},

(iii) {\em if ${D}_1>{D}_2$, then there is a  ${Q}>0$, such that ${V}_{rev}({Q})=0$.}

\noindent Considering the second case in above, the observations in Figure \ref{fig-VDVari} show that as $\rho=D_2/D_1$ increases,  magnitude of the corresponding $Q$ becomes larger. In fact, as $\rho \to \infty $, then $Q\to -\infty$.

\medskip

\noindent \underline{\textbf{Reversal potential ${V}_{rev}({Q})$ for    ${Q}=0$.}} 
For ${Q}=0$, one has ${V}_{rev}(0)=\theta\ln \frac{l}{r}$ from Theorem 4.2 in \cite{ML19} where $\theta=(D_2-D_1)/(D_2+D_1)$.
Therefore, 

(i) {\em if ${D}_1 = {D}_2$, then ${V}_{rev}(0)=0$,}

(ii) {\em if ${D}_1 \neq {D}_2$, then ${V}_{rev}(0)$ has the same sign as that of $\theta (l-r)$}.

\noindent Let us consider the case where $D_1 < D_2$ for a moment. 
In that case, ${V}_{rev}(0)$ has the same sign as that of $l-r$. This is reasonable, since for $V=0$ we have $|J_1| < |J_2|$ (since all but $J_k/D_k$ are independent of $D_k$ in \eqref{JonP}), and to help $|J_1|$ more than $|J_2|$ to get $J_1=J_2$ for zero current condition, one needs to increase $V$ when $l>r$ \big(and  decrease $V$ when $l<r$\big), and that is why $V_{rev}(0)>0$ for $l>r$ \big(and $V_{rev}(0)<0$ for $l<r$ \big). This is consistent with observations in Figure \ref{fig-VDVari} as well. The analysis for the other case, i.e., when ${D}_1 > {D}_2$ is similar.

\medskip

\noindent \underline{\textbf{Monotonicity of $\mathcal{V}_{rev}$ with respect to $\mathcal{Q}$.}}
It follows from Theorem 4.4 in \cite{ML19} that, for $\theta Q > 0$,  $\partial_{\mathcal{Q}}\mathcal{V}_{rev}$  has the same sign as that of $l-r$. This analytical result does not make conclusion   about the   case for  $\theta Q < 0$, though. The observations in Figures \ref{fig-Vrev} and \ref{fig-VDVari} show that result holds for any $\theta$ and $Q$. Thus we have 
\medskip

\noindent
{\em Conjecture: ${V}_{rev}$ is   increasing in ${Q}$ for $l>r$ and decreasing in ${Q}$ for $l<r$.}
\medskip

We remark that, in Figure \ref{fig-Vrev}, we take  $L = 20 mM,  R = 50 mM$, and $\mathcal{D}_1 = 1.334 \times 10^{-9} m^2/s $ and $\mathcal{D}_2 = 2.032 \times 10^{-9}m^2/s$ which are diffusion constants of Na$^+$ and Cl$^-$ respectively (see the solid line), and $\mathcal{D}_1 = 1.334 \times 10^{-9} m^2/s $ and $\mathcal{D}_2 = 0.792 \times 10^{-9}m^2/s$, where $\mathcal{D}_2$ is the diffusion constants of Ca$^{2+}$ (see the dashed line).

\begin{figure}[H]\label{fig-Vrev}
\centerline{\epsfxsize=3.0in
\epsfbox{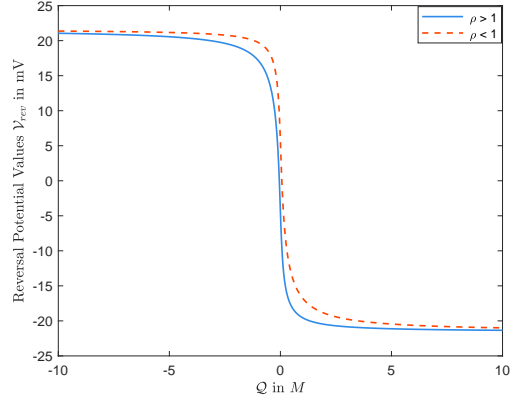} 
}
\caption{\em  $\mathcal{V}=\mathcal{V}_{rev}(\mathcal{Q})$ decreases when $L<R$, independent of values of diffusion constants.}
\label{fig-Vrev}
\end{figure}

\begin{figure}[H]\label{fig-VDVari}
\centerline{\epsfxsize=3.0in
\epsfbox{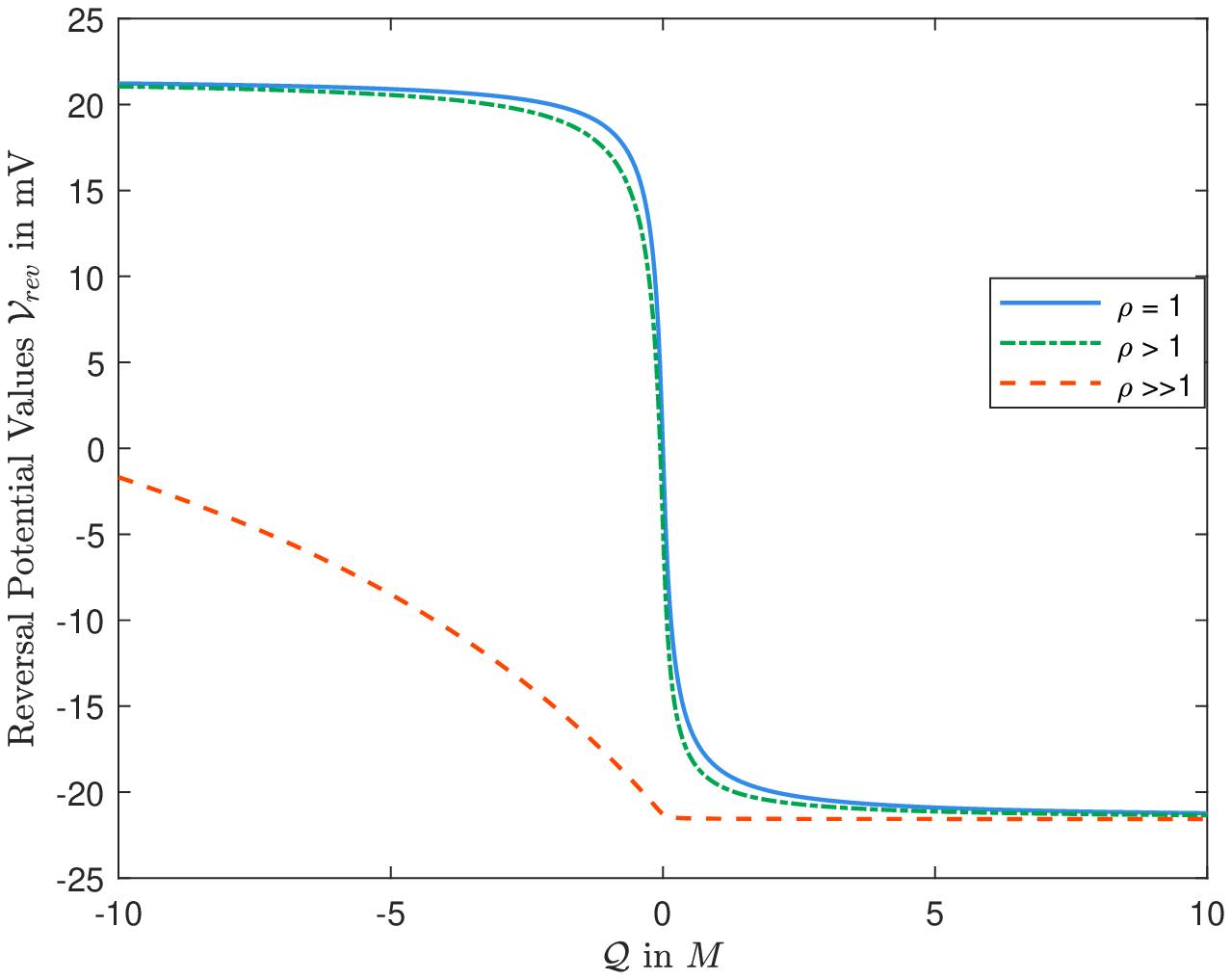} 
\epsfxsize=3.0in \epsfbox{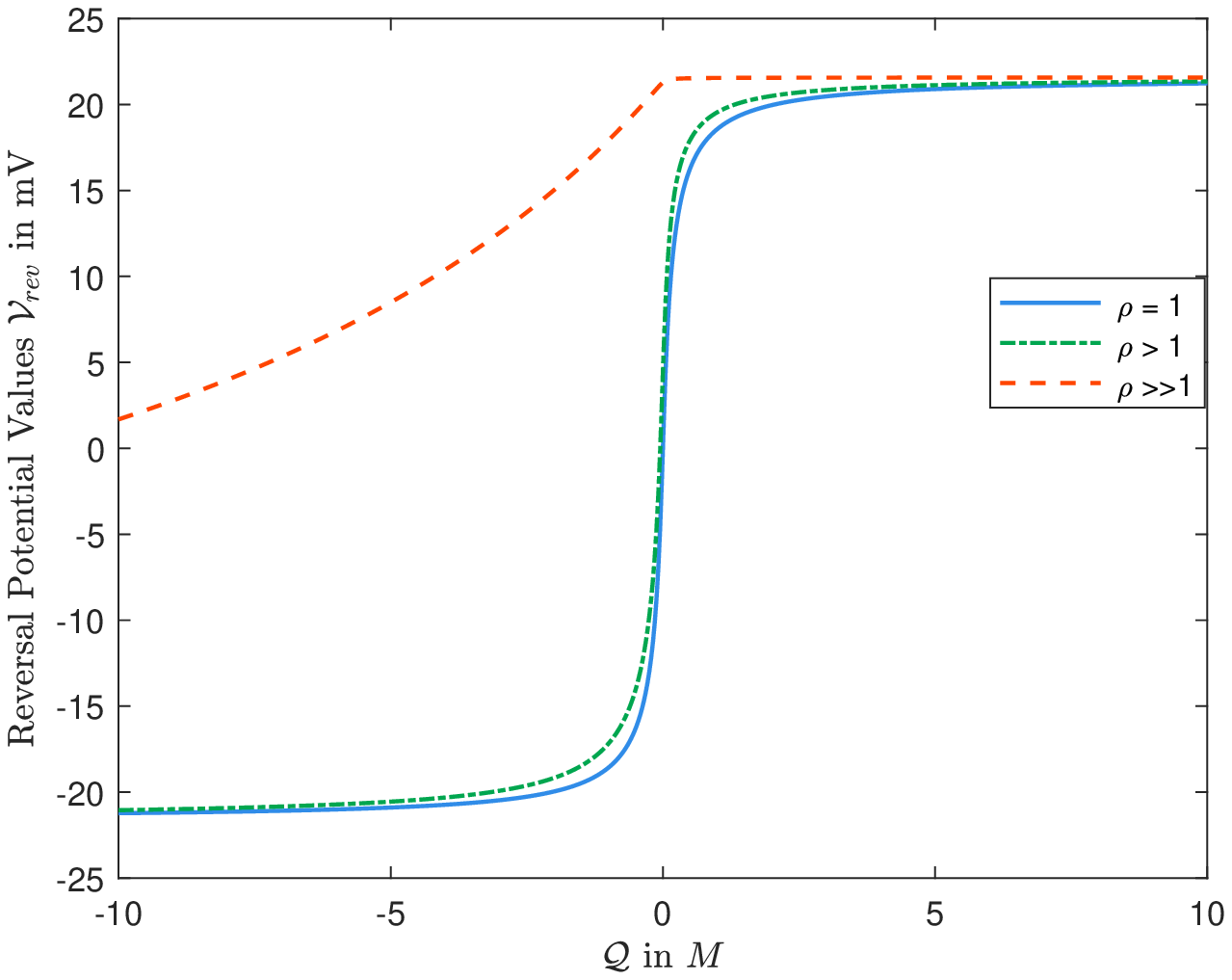} }
\caption{\em The function $\mathcal{V}=\mathcal{V}_{rev}(\mathcal{Q})$: The left panel for  $L=2 mM$ and $ R= 5 mM$; the right panel for $L= 5 mM$ and $R= 2 mM$.}
\label{fig-VDVari}
\end{figure}


\noindent \underline{\textbf{Dependence of $\mathcal{V}_{rev}$ on  $\rho = {D_2}/{D_1}$.}}
In terms of the dimensionless form of quantities (See Remark \ref{rem-dim1}), let us examine the dependence of  ${V}_{rev}$ on $\rho=D_2/{D_1}$ for effects of $D_1$ and $D_2$. It follows from Proposition 4.6 in \cite{ML19} that
\medskip

 {\em The reversal potential $ V_{rev}$ is increasing in $\rho$ if $l>r$ and  is decreasing in $\rho $ if $l<r$.}

\medskip

\noindent This feature reveals a fantastic aspect that is not intuitive immediately. 
Recall the equation \eqref{JonP}.
Given the boundary values and diffusion constants, the values one obtains for all terms in  (\ref{JonP}) except $J_k$ are independent of $D_k$ (\cite{Liu09}). 
The relation surely holds for the zero-current condition, i.e., $J_1=J_2$ with $V={V}_{rev}$. Now, let fix $D_1$ and increase $D_2$ (so $\rho$ is increasing). Then $|J_2|$ increases since all but $\frac{J_2}{D_2}$ in (\ref{JonP}) are independent of $D_2$. Consequently, to meet zero-current condition, we need to increase $|J_1|$. Intuitively increasing ${V}_{rev}$ seems to lead to an increase in $|J_1|$. This intuition agrees with the results only for $l>r$. In the other case, i.e., $l<r$, it is the exact opposite, though. That is, for $l<r$, it says, as $\rho$ increases, ${V}_{rev}$ decreases. This counterintuitive behavior could be explained by the fact that  $c_1(x)$ depends on ${V}_{rev}$, and reducing ${V}_{rev}$ could still  increase $|J_1|$.   
In fact,  $l<r$ will result in reducing ${V}_{rev}$, but $c_1(x)$ changes in a way that consequently increases $|J_1|$.

To illustrate the counterintuitive behavior, we provide a numerical result in Figure \ref{fig-Dincrease}. 
We choose $C_0, \hat{L}$  and $\mathcal{D}_1$ for Na$^+$, as it is mentioned in  Remark \ref{rem-dim2}.
 Now, suppose that $\mathcal{D}_2^1=0.792\times 10^{-9} m^2/s$, and consider the boundary concentrations  $L = 20 mM, R = 50 mM $ and $\mathcal{Q} = 1 M$. 
In this case $\mathcal{V}_{rev}=-16.7657 mV$ and $\mathcal{J}= -10.6184 MHz$. 
Now, if we increase $\mathcal{D}_2$ to $\mathcal{D}_2^2 = 2.032\times 10^{-9} m^2/s$ which is Cl$^-$ diffusion constant, then $\mathcal{V}_{rev} =-19.5527 mV$ and $\mathcal{J}= -11.3146 MHz$. These values make sense now, based on the above discussion.
Note that, we just pictured the middle part of the channel in Figure \ref{fig-Dincrease} since the sides are almost identical. One should notice that it is hard to realize, from Figure \ref{fig-Dincrease}, how $L<R$ will result in reducing $\mathcal{V}_{rev}$.
The complicated behavior discussed above convinces us that sometimes numerical results cannot necessarily help us if we do not know the whole truth from analytical results.

\begin{figure}[H]\label{fig-Dincrease}
\centerline{\epsfxsize=3.0in
\epsfbox{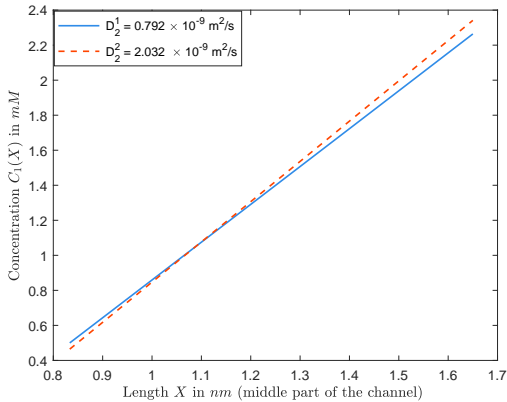} 
}
\caption{\em Graphs of $C_1(X)$ when $\mathcal{D}_1$ is fixed, but we increase $\mathcal{D}_2$. } 
\label{fig-Dincrease}
\end{figure}


\subsection{A comparison with Goldman-Hodgkin-Katz  equation for  $\mathcal{V}_{rev}$.}\label{Ef_of_Q}
In this section, we first recall the  GHK equation (\cite{Goldman,HK49}), which relates the reversal potential with the boundary concentrations and the permeabilities of the membrane to the ions.
If the membrane is permeable to only one ion, then that ion's Nernst potential is the reversal potential at which the electrical and chemical driving forces balance.  The GHK equation is a generalization of the Nernst equation
in which   the membrane is permeable to more than just one ion.   The derivation of GHK equation assumes that the electric field  across the lipid membrane is constant (or equivalently, the electric potential $\phi(x)$ is linear in $x$ in the PNP model).
Under the assumption, the I-V (current-voltage) relation is given by
\[I = V \sum_{k=1}^n z_k^2 D_k \dfrac{r_k - l_k e^{z_kV}}{1- e^{z_kV}}.\]
For the   case where $n=2$ and $z_1 = -z_2=1$,   the GHK  equation for the reversal potential    is  
\begin{align}\label{N-GHK}
{V}_{rev}^{GHK}(\rho)=  \ln \dfrac{r + \rho l}{l + \rho r}.
\end{align}

The assumption that the electric potential $\phi(x)$ is linear is not correct when applied to channels in proteins. That is because proteins have specialized structure and spatial distributions of permanent charge (acid and base side chains) and polarization (polar and nonpolar side chains). Experimental manipulations of the structure of channel proteins show that these properties control the biological function of the channel. The GHK equation does not contain variables to describe any of these properties and so cannot account for the biological functions they control.
 A linear  $\phi(x)$ is widely believed to make sense without channel structure presumably, in particular, $Q_0=0$. However, this is not correct, either.  It follows from \eqref{VonQf} for $Q_0=0$ that the zeroth order in $\varepsilon$ approximation of the  reversal potential in this case is,
 \begin{equation}\label{VRevWOQ}
{V}_{rev}(0,\rho) =  \dfrac{\rho -1}{\rho + 1} \ln \dfrac{l}{r}.
 \end{equation}
  Figure \ref{fig-VvsGHK}  compares ${V}_{rev}(0,\rho)$ in \eqref{VRevWOQ}  with ${V}_{rev}^{GHK}$ from the GHK-equation in (\ref{N-GHK}).   It can be seen, from the left panel that when $l$ and $r$ are close (for example $L=C_0l= 20 mM, R=C_0 r = 50mM$), then the two curves have almost the same behavior. However, when we reduce $L$ from $ 20mM$ to $1 mM$, then the right panel shows a significant difference between the two graphs.


\begin{figure}[H]\label{fig-VvsGHK}
\centerline{\epsfxsize=3.0in
\epsfbox{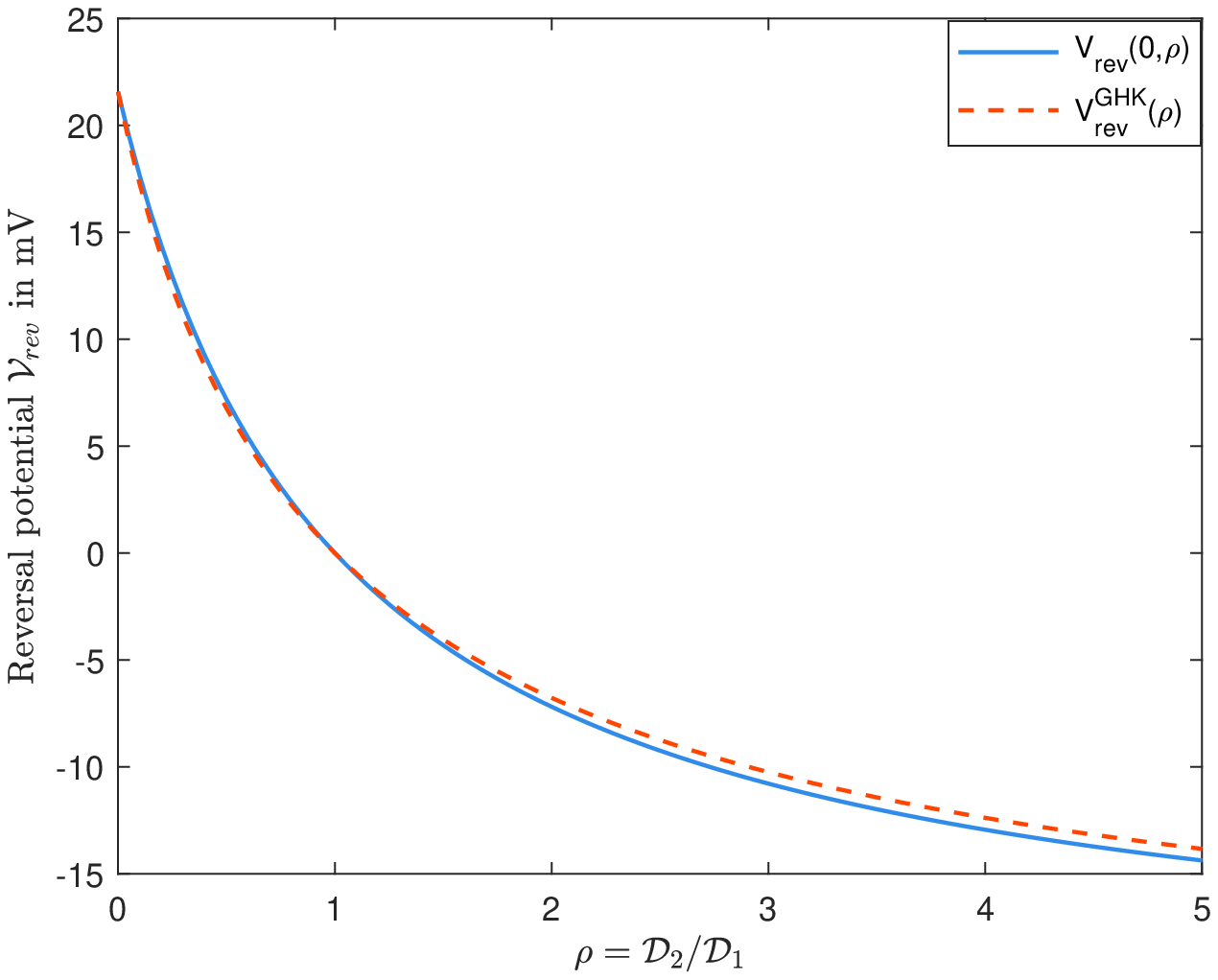}
\epsfxsize=3.0in 
\epsfbox{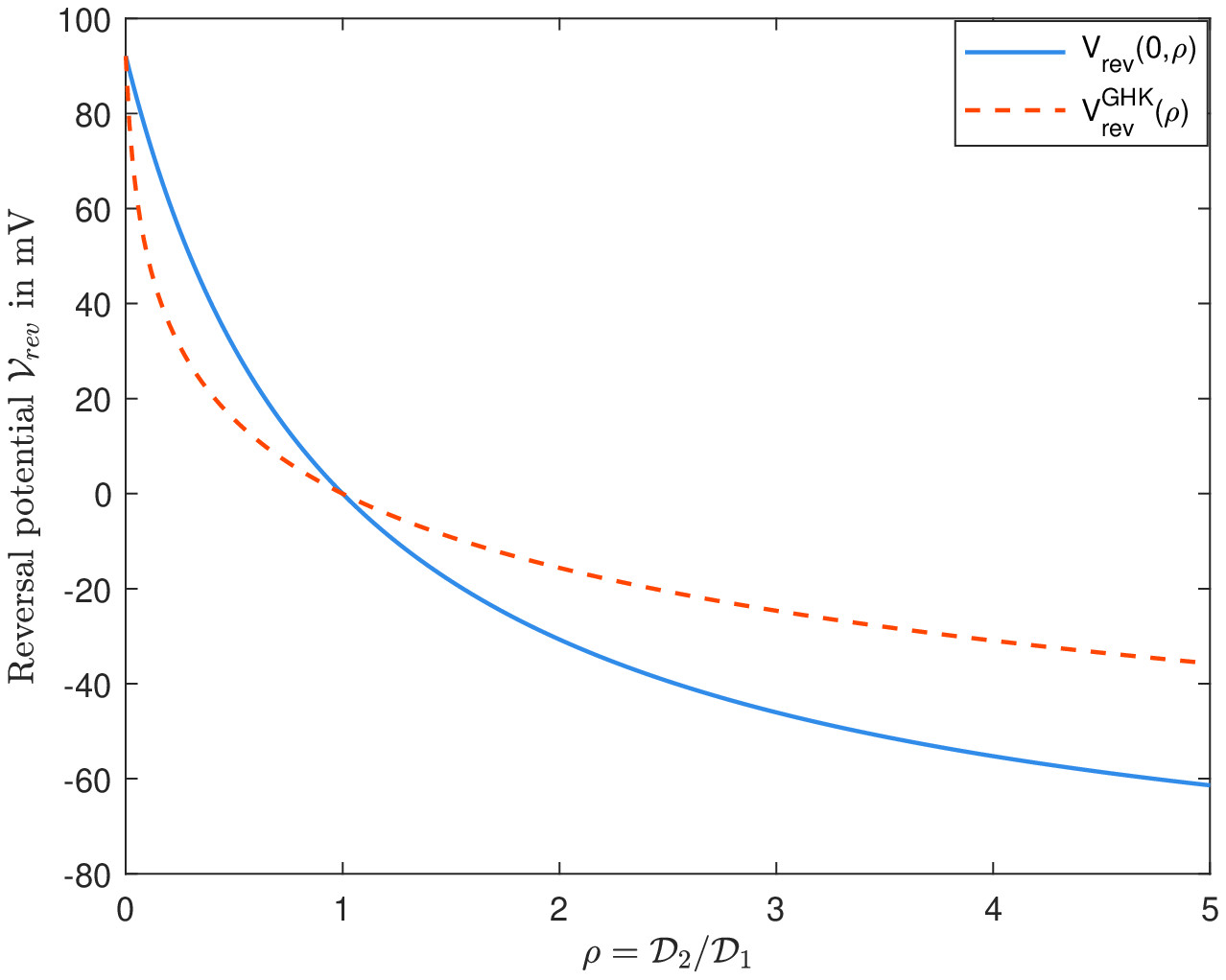} }
\caption{\em  $\mathcal{V}_{rev}(Q =0, \rho)$ vs $\mathcal{V}_{rev}^{GHK}(\rho)$: The left panel for  $L=20 mM$ and $R= 50 mM$; the right panel for  $L= 1 mM$ and $R= 50 mM$.} 
\label{fig-VvsGHK}
\end{figure}

In Figure \ref{fig-VvsRhoQ}, we arrange a simple numerical result for the case where $\mathcal{Q} \neq 0$ to compare the graphs of $\mathcal{V}_{rev}(\mathcal{Q}, \rho)$, obtained from \eqref{VonQf},  for various values of permanent charge $\mathcal{Q}$. We consider $ L = 20 mM,  R = 50 mM $, and $0 < \rho <5$ for some values of $\mathcal{Q}$, i.e., $\mathcal{Q}=0 M, 1 M,10 M$.

\begin{figure}[H]\label{fig-VvsRhoQ}
\centerline{\epsfxsize=3.0in
\epsfbox{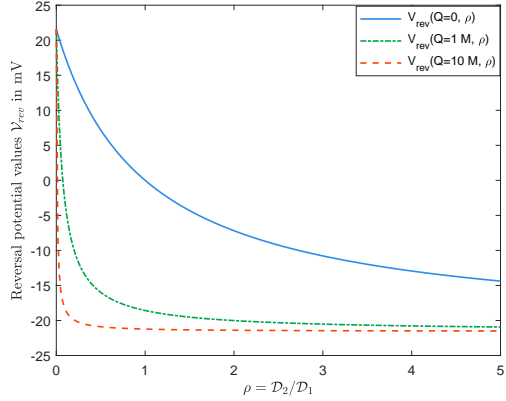} 
}
\caption{\em  $\mathcal{V}_{rev}(\mathcal{Q} , \rho)$ with various values of permanent charges.  }
\label{fig-VvsRhoQ}
\end{figure}
\subsection{Profiles of relevant physical quantities.}\label{sec-PhysQuan} It follows from \eqref{G1G2Sys} and \eqref{G} that for any given $Q$, once a solution  $(A,V)$ are determined, all the other unknowns can be  determined. 
We consider the dimensional form of quantities, and fix $(\mathcal{Q},L,R, \mathcal{D}_1, \mathcal{D}_2)$ to numerically investigate the behavior of $C_k(X)$ and $\Phi(X)$ throughout the channel. Figures \ref{fig-C10} and \ref{fig-Mu10} graph the cases with  small permanent charge $\mathcal{Q} =0.0001 M = 0.1 mM$ when  $L= 20 mM,  R = 50 mM$, $\mathcal{D}_1=1.334 \times 10^{-9} m^2/s,$ and $ \mathcal{D}_2=2.032 \times 10^{-9} m^2/s$. In this case, we obtain $\mathcal{J}= -72.7387 MHz$ and $\mathcal{V}_{rev} = -4.4820 mV$.
\begin{figure}[H]\label{fig-C10}
\centerline{\epsfxsize=3.0in \epsfbox{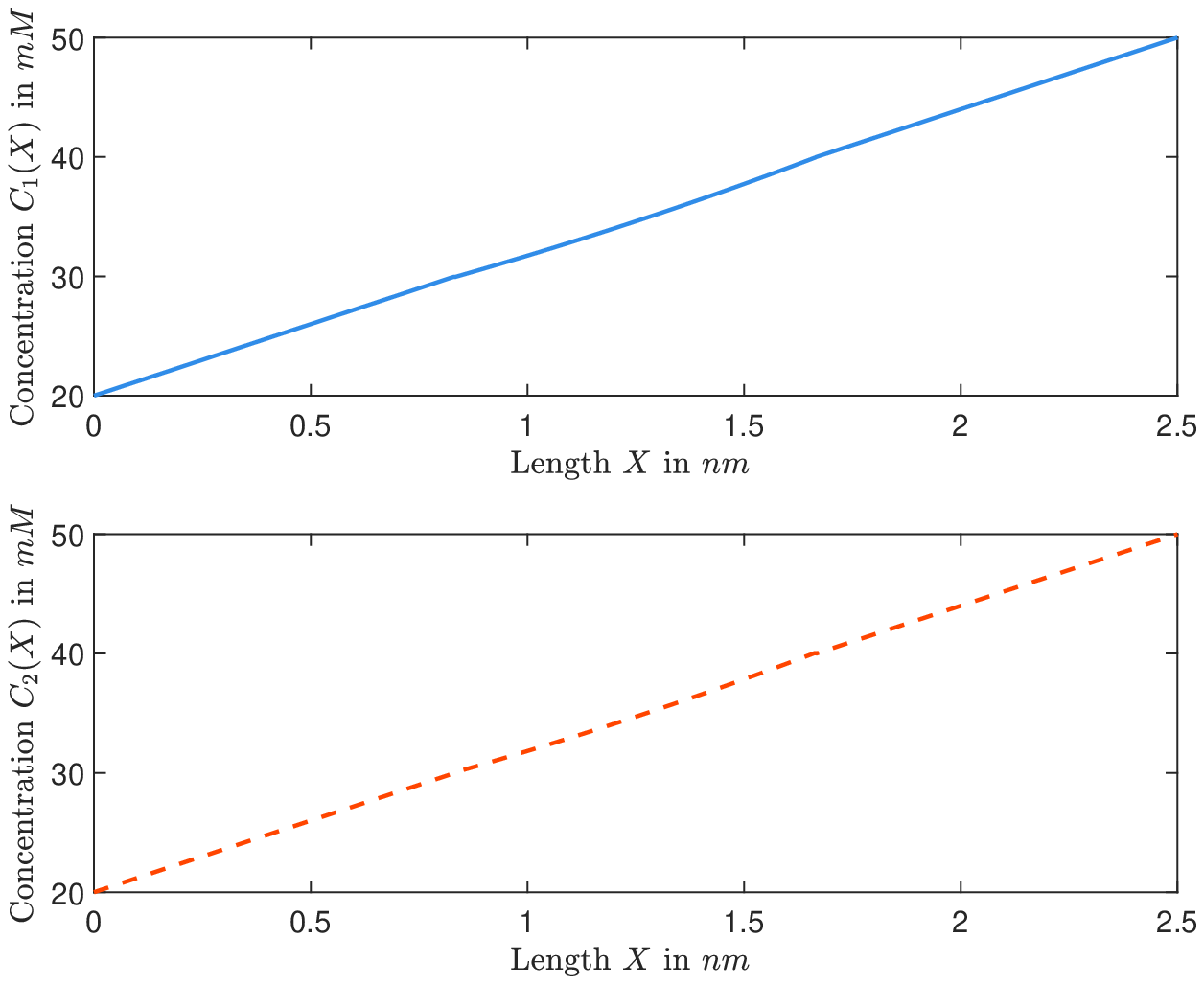} 
\epsfxsize=3.0in \epsfbox{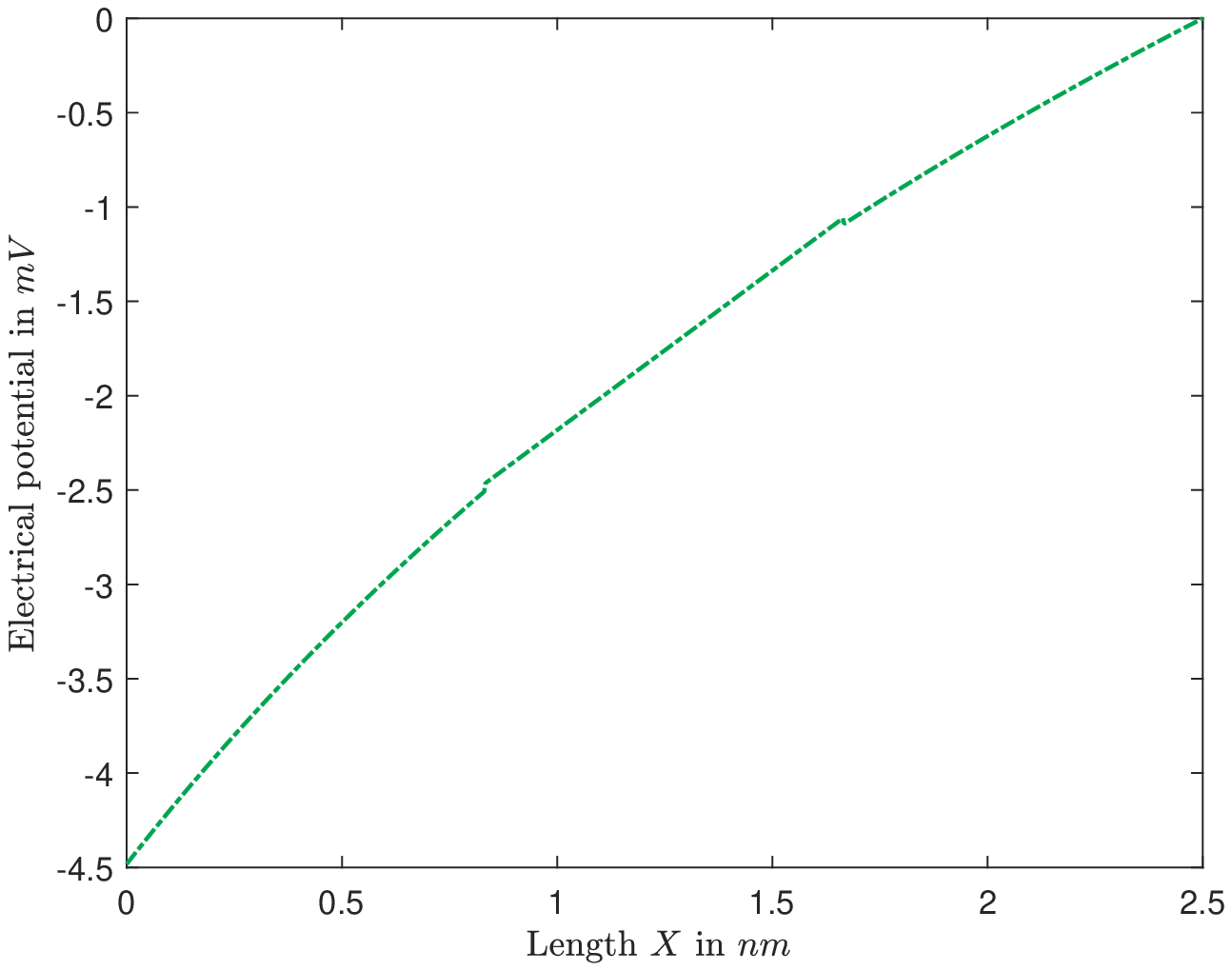}
}
\caption{\em The functions $C_k(X)$ (left) and  $\Phi(X)$ (right) with $\mathcal{Q}= 0.1 mM$.}
\label{fig-C10}
\end{figure}


Furthermore, Figures \ref{fig-C11} and Figure\ref{fig-Mu11} show graphs of concentrations, electrical potential, and  electrochemical potentials versus $X$, where $L =20 mM, R= 50 mM, Q =2 M,$ and diffusion constants are the same as previous one.
In this case, we obtain $\mathcal{J}= -11.3146 MHz$ and $\mathcal{V}_{rev} = -19.5527 mV$.
\begin{figure}[H]\label{fig-Mu10}
\centerline{\epsfxsize=3.0in \epsfbox{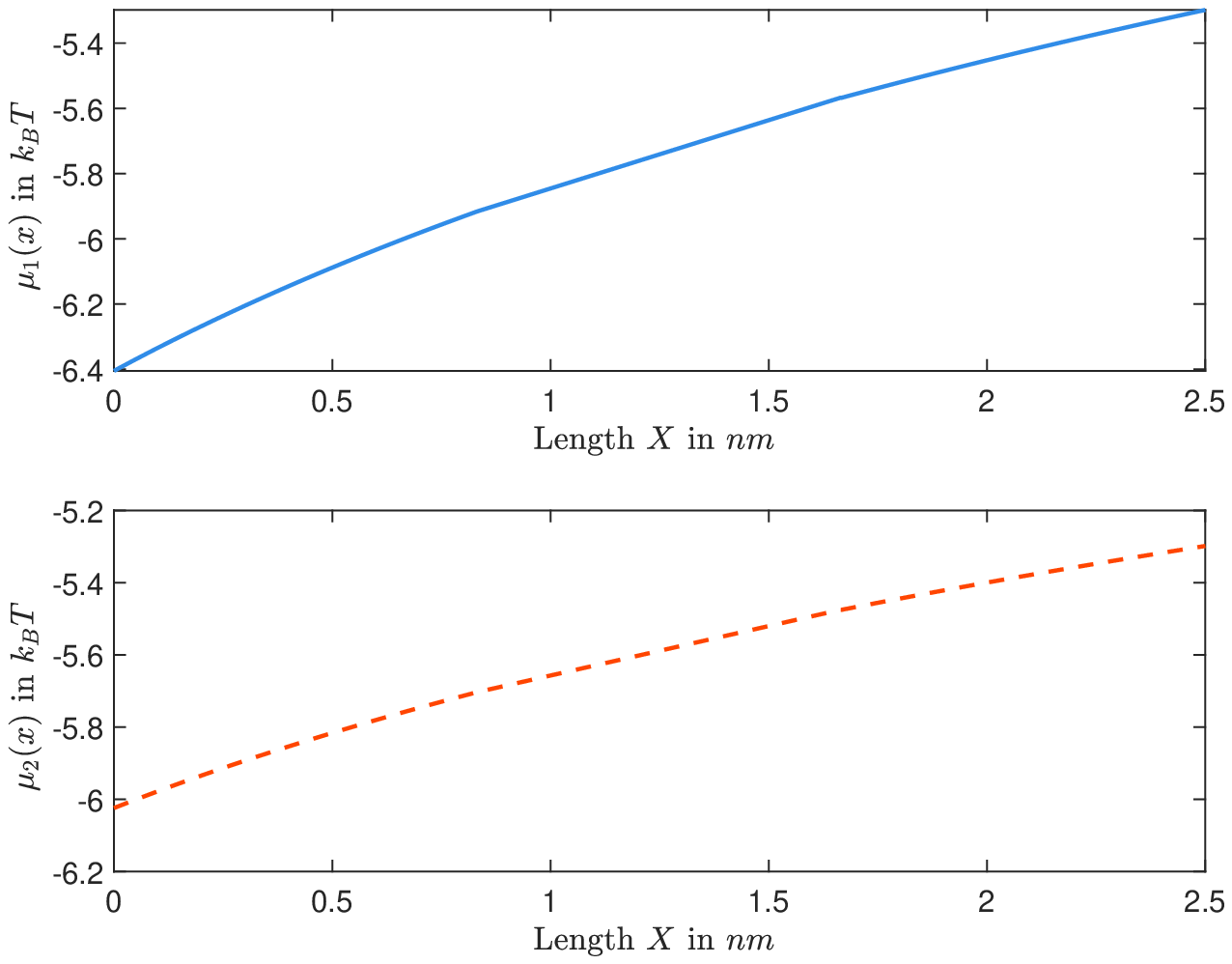} }
\caption{\em The functions $\mu_1(X)$ and $\mu_2(X)$ are increasing for $L<R$ and  $\mathcal{Q}= 0.1 mM$.}
\label{fig-Mu10}
\end{figure}


\begin{figure}[H]\label{fig-C11}
\centerline{\epsfxsize=3.0in \epsfbox{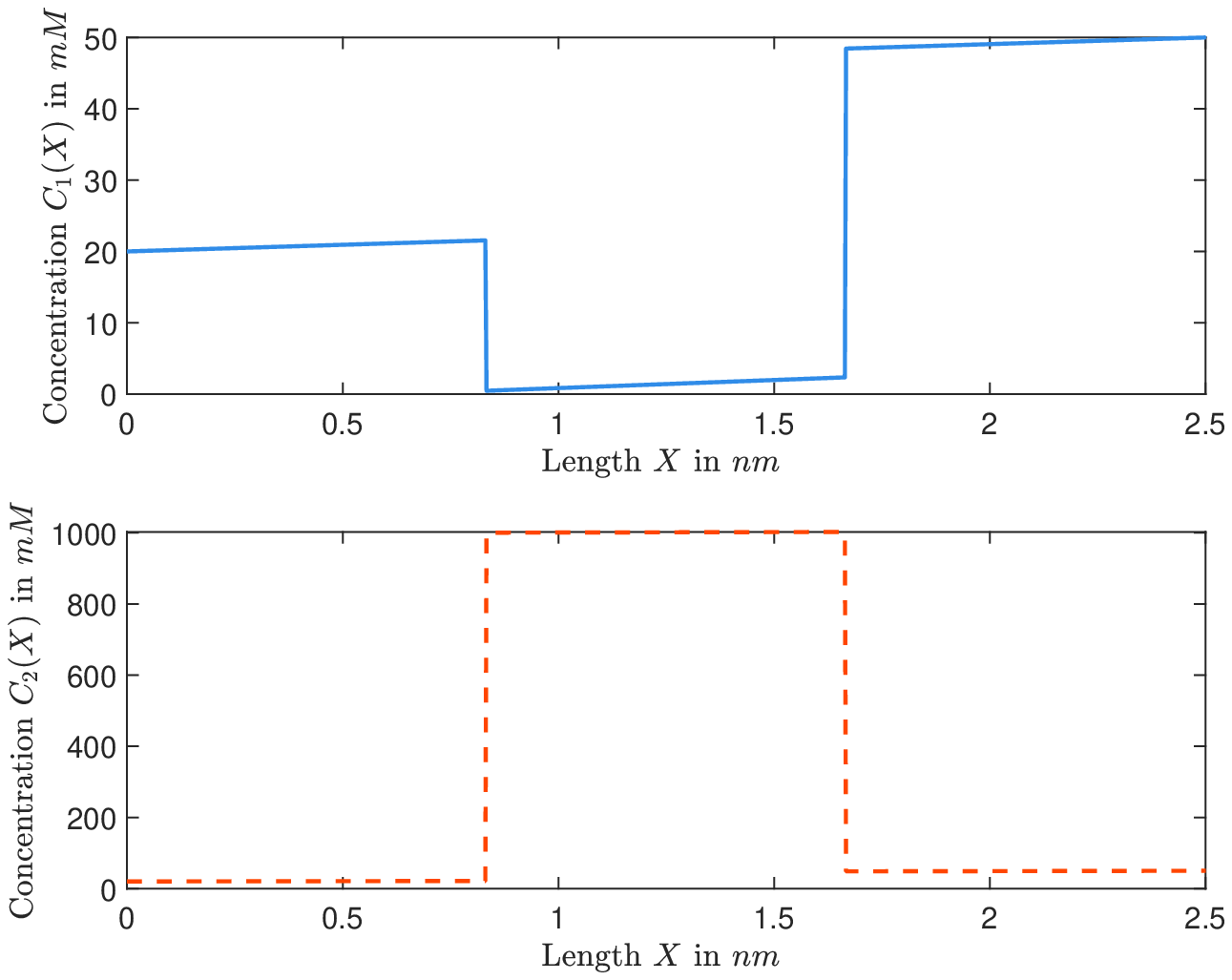} 
\epsfxsize=3.0in \epsfbox{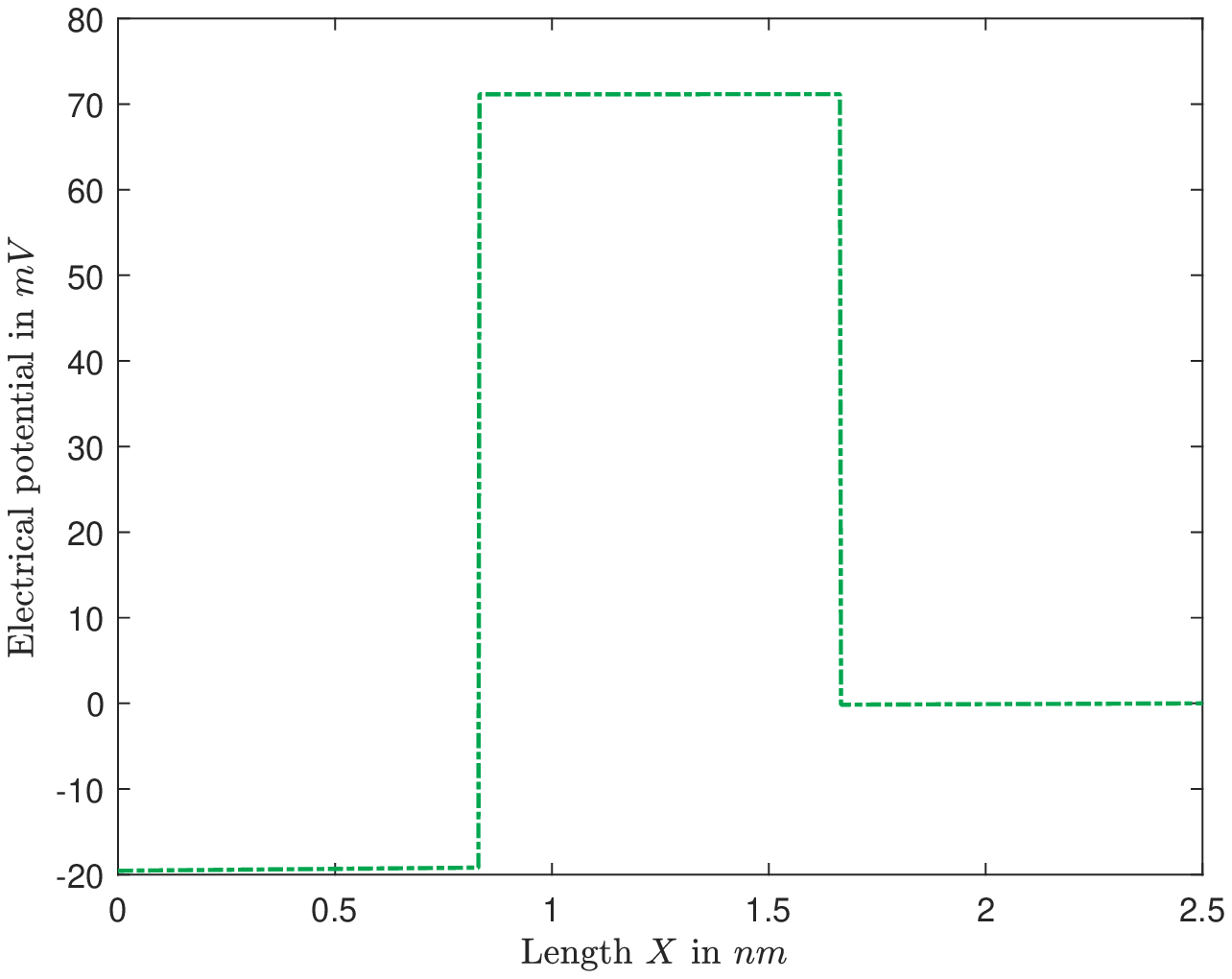}
}
\caption{\em The functions $C_1(X)$ and $C_2(X)$ (left) and the function $\Phi(X)$ (right) for $\mathcal{Q}= 1 M$.}
\label{fig-C11}
\end{figure}
\begin{figure}[H]\label{fig-Mu11}
\centerline{\epsfxsize=3.0in \epsfbox{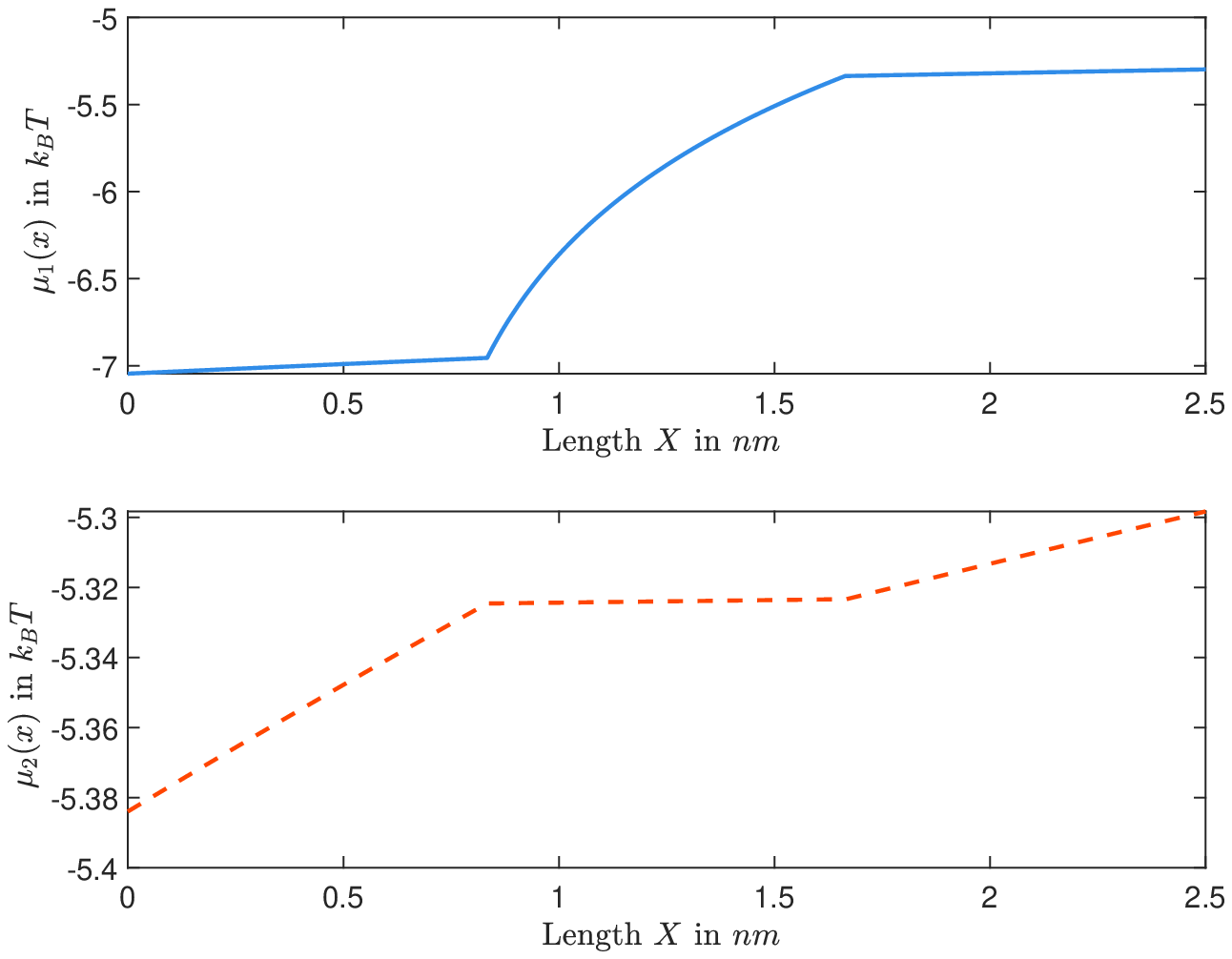} }
\caption{\em The functions $\mu_1(X)$ and $\mu_2(X)$ are increasing for $L<R$ and  $\mathcal{Q} = 1 M$.}
\label{fig-Mu11}
\end{figure}

\section{The current-voltage and current-permanent charge behaviors. } \label{sec-CurVol}
\setcounter{equation}{0}

Ionic movements across membranes lead to the generation of electrical currents. The current carried by ions can be examined through {\em current-voltage} relation or I-V curve. 
In such a case, voltage refers to the voltage across a membrane potential, and current is the flow of charged ions through channels in the membrane. Another important data is {\em current-permanent charge}, i.e. I-Q relation.
Dependence of current on membrane potentials and permanent charge is investigated in this section for arbitrary values of diffusion constants.  

To derive the I-V and I-Q relations, we rely on \cite{EL} where the authors  showed that the set of nonlinear algebraic equations is equivalent to one nonlinear equation depending on $A$, defined in \eqref{A}.
All other variables can be obtained from the special variable $A$. It is crucial to realize that this is a specific result, not possible in many cases. One can only imagine that the resulting simplification produces controllable and robust behavior that proved useful as evolution designed and refined protein channels. The reduction allowed by this composite variable can be postulated to be a ``design principle" of channel construction, in technological (engineering) language, or an evolutionary adaptation, in biological language. In particular, the current $I$ can be explicitly expressed in terms of boundary conditions, permanent charge, diffusion constants, and transmembrane potential in the special case that allows the definition of $A$. In what follows, we derive flux and current equations -- when diffusion constants are involved as well -- in terms of boundary concentrations, membrane potential, and permanent charge. The I-V, I-Q, J-V, and J-Q relations are investigated afterward in section \ref{sec-IVQ}.

\subsection{Reduced flux and current equations.}\label{sec-FlCu}

It was shown in \cite{EL} that  the singular orbits of BVP \eqref{1dPNPdim} can be reduced to the algebraic equation 
\begin{equation}\label{Redu-A}
\eta \ln\dfrac{S_b -\eta}{S_a - \eta} -N =0,
\end{equation}
where $B= l-A+r$, and $S_a, S_b$ and $N$ were defined in \eqref{Nab}, and,
 \begin{equation}\label{JD}
\eta = Q_0 - \dfrac{Q_0}{\ln\frac{Bl}{Ar}}\Big( V + \ln\dfrac{l(S_b-Q_0)}{r(S_a-Q_0)}\Big) + \dfrac{N}{\ln\frac{Bl}{Ar}}.
\end{equation}
Once $A$ is solved from \eqref{Redu-A},  we can obtain  the flux densities and current equations as follows,
\begin{equation}\label{IJ}
\begin{aligned}
J_k := J_k(V,l,r,D_1,D_2) =& 3D_k(l-A)\Big(1+(-1)^k\frac{\eta}{Q_0}\Big), \quad \text{for} ~k=1,2,\\
I := I(V,l,r,D_1,D_2)  = & J_1 - J_2 =  3(l-A)\Big(D_1 -D_2-\frac{\eta}{Q_0}(D_1+D_2)\Big).
\end{aligned}
\end{equation}
For any given $(l,r,D_1,D_2,Q_0,V)$   there exists a solution for the flux $J$ and current $I$.  The numerical results in the next section give us more information on ``current-voltage" and ``current-permanent charge" relations.  

\subsection{Current-voltage and current-permanent charge relations.}\label{sec-IVQ}
%
%
%
%

\noindent \underline{\bf Dependence of current on diffusion constants.}
Now we reveal a remarkable feature of the theoretical results that is not intuitive.  Suppose that $(l,r, Q)$ is given ($V$ is still free!). It follows from \eqref{Nab} for the definition of $N$ that there exists an $A$ so that $N=0$. It consequently follows from \eqref{JD} that, if  
$V= \ln \dfrac{B(S_a - Q_0)}{A(S_b- Q_0)}$, then $\eta = 0$.  
Therefore, from (\ref{IJ}), $I = 3(l-A)(D_1-D_2)$, which implies, 
\medskip

{\em For some fixed parameters  $(l, r, V, Q)$,  the sign of I depends on the sign of $D_1-D_2$.}

  \medskip

\noindent\underline{\textbf{Current-voltage relations or I-V curves.}}  
Figure \ref{fig-IV1} is a numerical simulation from \eqref{Redu-A} and \eqref{JD} of the I-V  curves for some values of $\mathcal{Q}$ with $\mathcal{D}_1=1.334 \times 10^{-9} m^2/s$ and $ \mathcal{D}_2=2.032 \times 10^{-9} m^2/s$.
One may conjecture, incorrectly based on the numerical observations,  that the value of current $\mathcal{I}$, obtained from \eqref{IJ}, is unique for any $\mathcal{V}$ and $\mathcal{Q}$. However, this is incorrect, in general, as mentioned in previous section.  The existence of a unique solution is important since the opening and closing properties of channels might be thought to arise from non-unique solutions (\cite{Eis96, Eis96(2)}). 

\begin{figure}[H]\label{fig-IV1}
\centerline{\epsfxsize=3.0in
\epsfbox{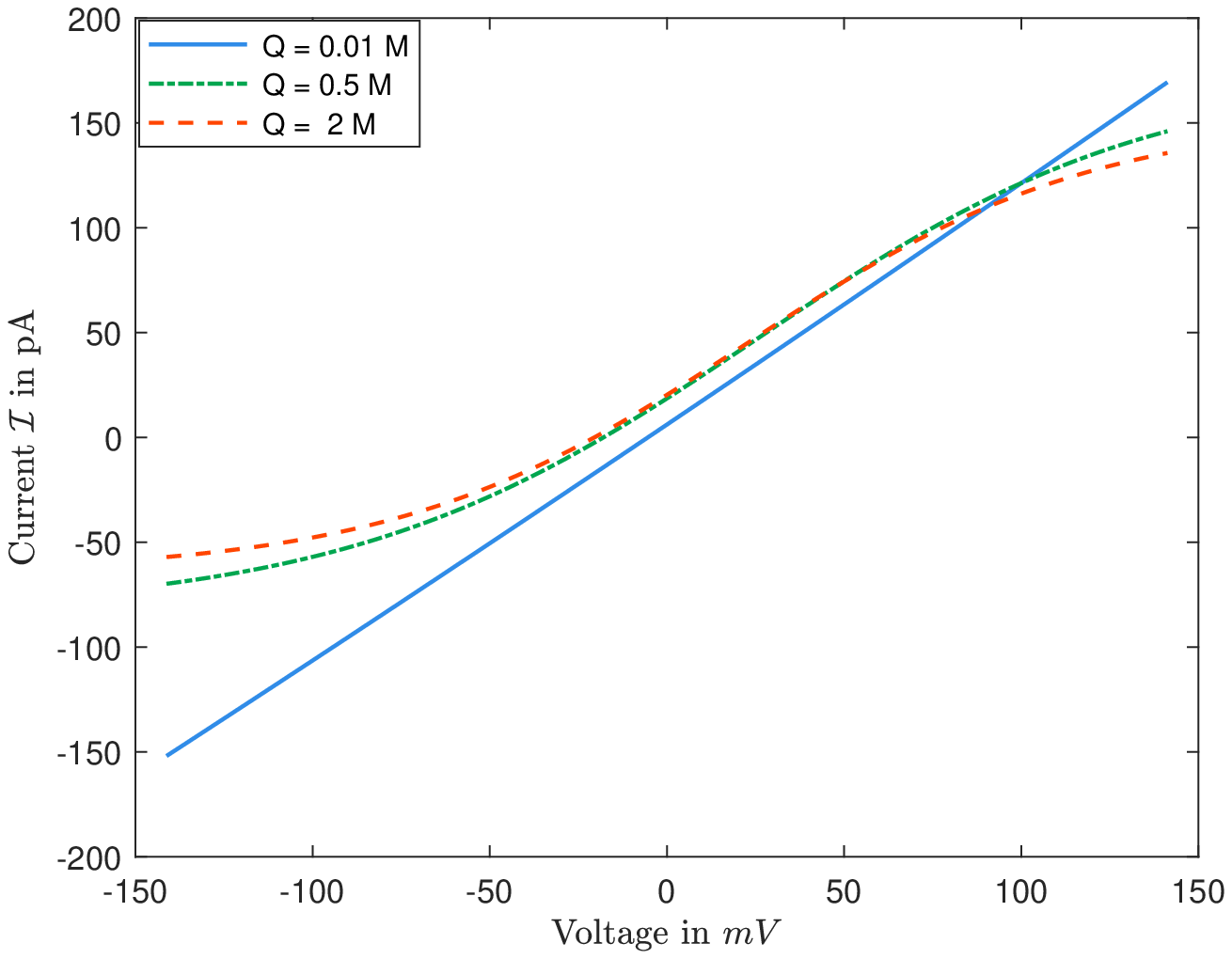} 
}
\caption{\em The function $\cal{I}=\cal{I}(\cal{V})$ for  $L=20 mM $ and $R=50 mM$. }
\label{fig-IV1}
\end{figure}

\noindent \underline{\textbf{Current-permanent charge relations or I-Q curves.}} Note that the equation \eqref{JonP}, in dimensional form is,
$$
\mathcal{J}_k \int_0^{\hat{L}}\dfrac{k_BT}{\mathcal{D}_kA(X)C_k(X)}dX= \mu_k(0) -\mu_k(\hat{L}), \quad k=1,2.
$$
The sign of $\mathcal{J}_k$ is determined by the boundary conditions, independent of the permanent charge $\mathcal{Q}$. Nevertheless, as seen in Figure \ref{fig-IQ2}, the magnitudes of $\mathcal{J}_k$'s, and consequently, the sign and the size of the current $\mathcal{I}$ do depend on the permanent charge $\mathcal{Q}$ in general. 

The other interesting counterintuitive observation for I-Q relations is the different ways current $\mathcal{I}$ behaves with respect to permanent charges $\mathcal{Q}$, for various values of $\mathcal{V}$. 
It follows from numerical observations (see Figure \ref{fig-IQ2}, for example) that there exists some $\mathcal{V}^*(\mathcal{D}_1, \mathcal{D}_2)>0$, independent of $L,R$, such that for any $\mathcal{V}$ where $\mathcal{V}>\mathcal{V}^*$, there exists a unique maximum for the current, $\mathcal{I}_{max}$, that corresponds to $\mathcal{V}$.  Likewise, there exists some $\mathcal{V}_*(\mathcal{D}_1, \mathcal{D}_2)<0$, independent of $L,R$, such that for any $\mathcal{V}$ where $\mathcal{V}<\mathcal{V}_*$, there exists a unique minimum for the current, $\mathcal{I}_{min}$, that corresponds to $\mathcal{V}$. 

\begin{figure}[H]\label{fig-IQ2}
\centerline{\epsfxsize=3.0in
\epsfbox{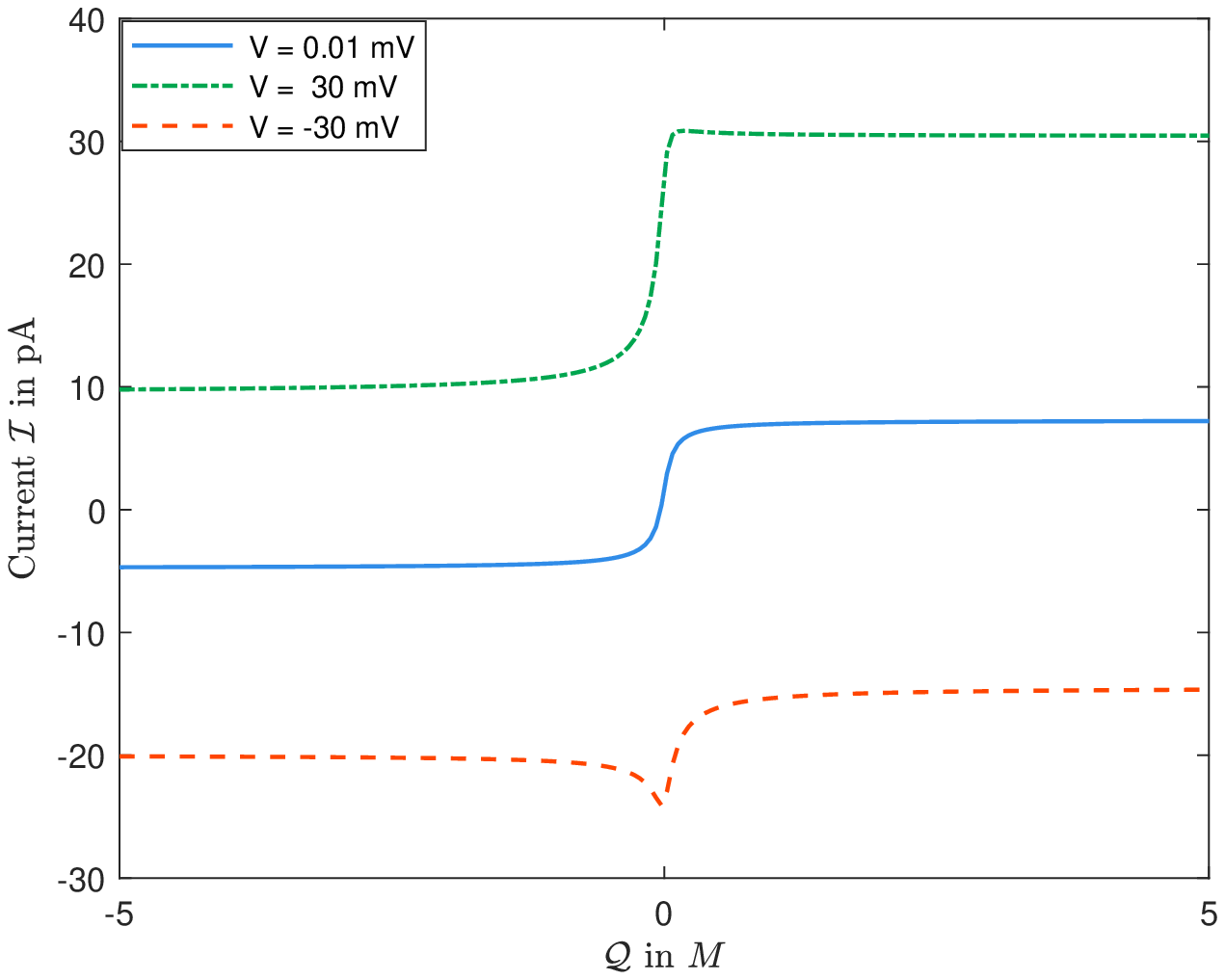} 
\epsfxsize=3.0in
\epsfbox{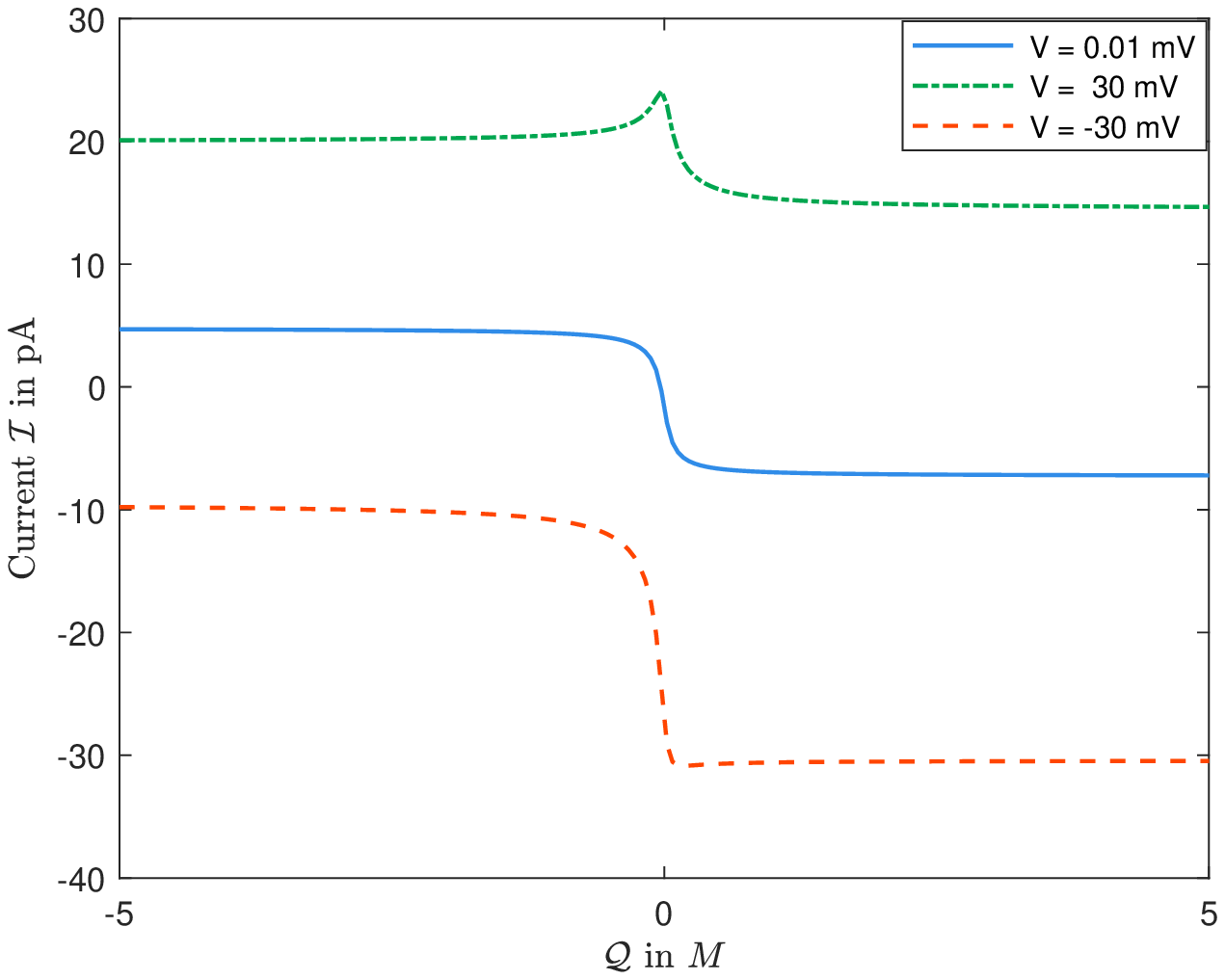}}
\caption{\em The function $\mathcal{I}=\mathcal{I}(\mathcal{Q})$ with $\mathcal{D}_1 < \mathcal{D}_2$: The left panel for $L=20 mM$ and $R=30 mM$;  the right panel for $L= 30 mM$ and  $R= 20 mM$.}
\label{fig-IQ2}
\end{figure}

Besides, we claim based on numerical observations (not proven though) that there exists $\hat{\mathcal{V}}(\mathcal{D}_1, \mathcal{D}_2)=\min \{\mathcal{V}^*,|\mathcal{V}_*|\}$, such that

(i) for any given $\mathcal{V}$ where $|\mathcal{V}|>\hat{\mathcal{V}}$, the corresponding current $\mathcal{I}$ is non-monotonic in $\mathcal{Q}$, but 

(ii) for any  $\mathcal{V}$ where $|\mathcal{V}|<\hat{\mathcal{V}}$, the corresponding current $\mathcal{I}$ is monotonic in $\mathcal{Q}$. 

\noindent In particular, it can  be seen, in section \ref{sec-ZPotCur}, that current is monotonic in $\mathcal{Q}$ for $\mathcal{V}=0$. 
In the end, we would like to mention that the diffusion constants affect the values $V^*$ and $V_*$ above.

\subsection{Flux-voltage and flux-permanent charge relations.}\label{sec-JVQ}
%
%

\noindent \underline{\textbf{Flux-voltage relations via J-V curves.}}  Now, we study flux and membrane potential relations, i.e., J-V curves, for given values of permanent charges $\mathcal{Q}$. In all cells, maintenance of cell volume is essential for survival. The membrane potential is likely to be a key regulator of this process in many cells.
The membrane potential, feeds into the cell volume, controls mechanism by changing the “driving force” for ionic fluxes. 

 The flux $\mathcal{J}$ depends on concentrations, diffusion constants, and the electrical potential across the membrane.
Without permanent charges, i.e., when $\mathcal{Q}=0$, the flux of one ion species is independent of the other quantities, based on the classical PNP models for dilute mixtures (as is well-known). With the presence of a permanent charge, as expected, the classical PNP model shows the dependence of the flux of one ion species on the other ion species \cite{JLZ15}. In particular, the effects of permanent charges on ionic flows could be intricate, depending on the interactions between boundary conditions and the channel geometry. 

\medskip

\noindent \underline{\textbf{Flux-permanent charge relations via J-Q curves.}} Numerical observations (see Figure \ref{fig-JQ1}) express  an interesting feature that is not easy to prove theoretically.
We consider the case where $l<r$. One has,

{\em (i) ~if $V < - \ln \frac{l}{r}$, then $J_1 <0$, and if $V > - \ln \frac{l}{r}$, then $J_1 >0$;

(ii) if $V <  \ln \frac{l}{r}$, then $J_2 > 0$, and if $V >  \ln \frac{l}{r}$, then $J_2 <0$.

  A similar discussion holds for the other case when $l>r$. }

We discuss the case (i) for $J_1$ on the left panel  in Figure \ref{fig-JQ1}. Note that if one takes $l = 0.002$ and $r = 0.003$ in dimensionless forms (i.e., $L = C_0l = 20mV,~R=C_0r =30mV$ in dimensional forms), then $\ln \frac{l}{r} = -0.4055$. Hence, $V=0 < - \ln \frac{l}{r}= 0.4055$ in this case, and $J_1 <0$ as stated in (i). However, if one considers $V=0.5$ (i.e., in dimensional form $\mathcal{V} = 11.7696 mV$), then $V= 0.5 > - \ln \frac{l}{r}$, and $J_1 >0$. 
\vspace*{-.1in}
\begin{figure}[H]\label{fig-JQ1}
\centerline{\epsfxsize=3.0in
\epsfbox{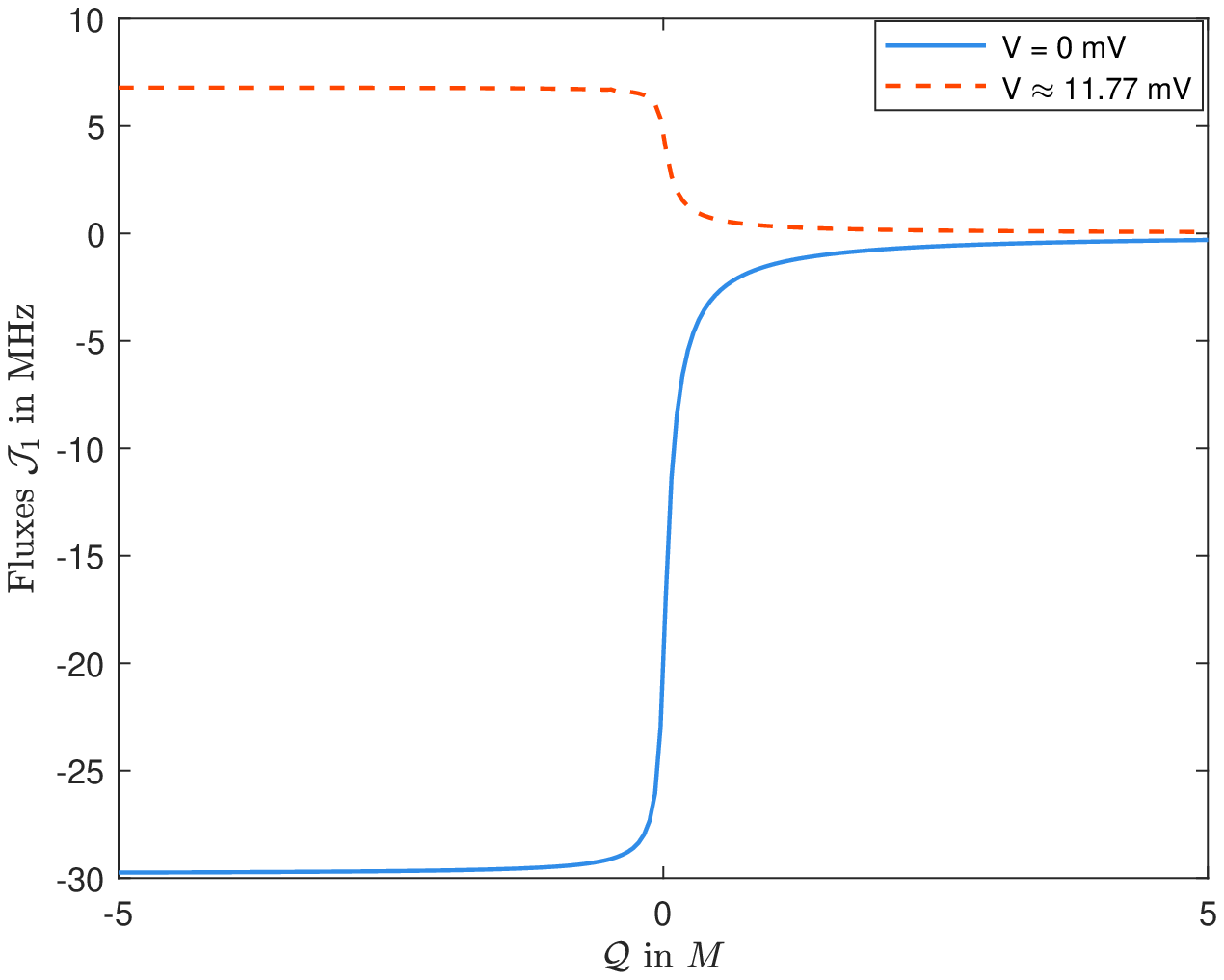} 
\epsfxsize=3.0in
\epsfbox{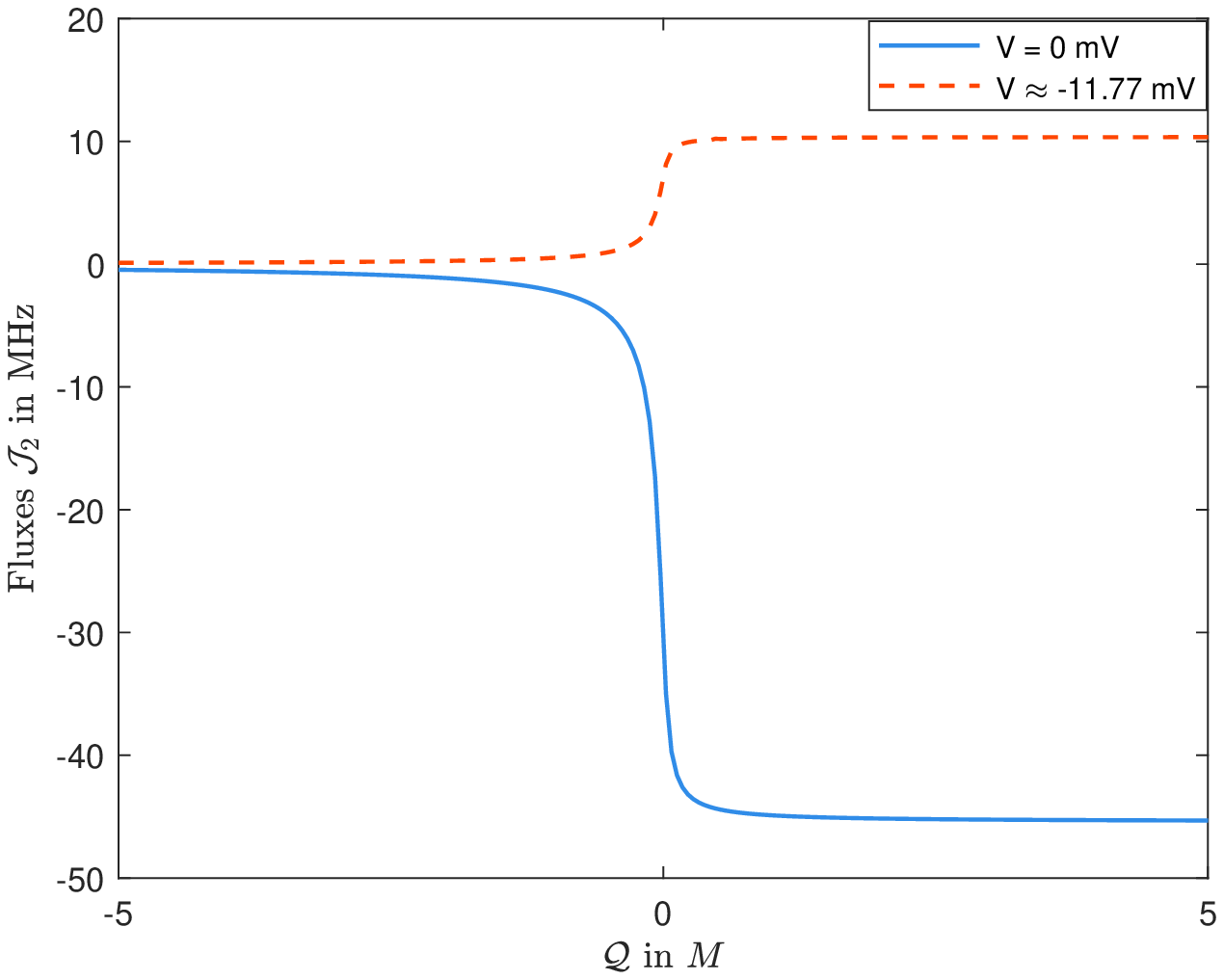}}
\caption{\em The functions $\mathcal{J}_1=\mathcal{J}_1(\mathcal{Q})$ (the left panel), and  $\mathcal{J}_2=\mathcal{J}_2(\mathcal{Q})$ (the right panel), with $L= 20 mM$ and $R= 30 mM$. }\vspace*{-.03in}
\label{fig-JQ1}
\end{figure}

\subsection{Zero-voltage current  and zero-voltage fluxes.}\label{sec-ZPotCur}
The different permeability of the membrane determines the zero membrane potential (voltage) to different types of ions, as well as the concentrations of the ions, the permanent charge, and the shape of the channel. The current $I(V)$, and the fluxes $J_k(V)$, for $k=1,2$, are called {\em zero-potential current} and {\em zero-potential fluxes} respectively  when ${V}=0$. 
For any given value of membrane potential $V$, the formulas for the current $I(V)$, for  small and large values of permanent charge $Q$, are provided in \cite{JLZ15} and \cite{ZEL} respectively. It follows from  \cite{JLZ15} that for small values of $Q$, applying $V=0$, zero-potential current $I^s(0)$, and zero-potential fluxes $J_k^s(0)$,(in dimensionless forms as mentioned in Remark \ref{rem-dim1}) are, \vspace*{-.1in}
\begin{equation}\label{V0I-smallQ}
\begin{aligned}
I^s(0) =& (l-r)\big(D_1 -D_2\big) - \dfrac{(l-r)^2}{(2l + r)(l + 2 r)\ln \frac{l}{r}}Q +O(Q^2),\\
{J}^s_{k}(0) =& (l-r)D_1+(-1)^k \dfrac{3(l-r)^2D_k}{2(2l + r)(l + 2 r)\ln \frac{l}{r}}Q+O(Q^2), \quad k=1,2.
\end{aligned}
\end{equation}
\vspace*{-.04in}
Furthermore, for large positive values of $Q=2Q_0$, with $\nu = \frac{1}{Q_0}$ (where $\nu$ is small), it follows from \cite{ZEL} that zero-potential current $I^l(0)$ and zero-potential fluxes ${J}^l_{k}(0)$ are,
\begin{equation}\label{V0I-LargeQ}
\begin{aligned}
I^l(0) = &  -  6D_2 \sqrt{lr}\dfrac{(\sqrt{l}- \sqrt{r})}{\sqrt{l} +  \sqrt{r}}+  \dfrac{ 3}{2}D_1 \Big(\dfrac{l +  r}{\sqrt{l} + \sqrt{r}} \Big)^2 (l -r) \nu\\
 &+\dfrac{3(l+r)D_2}{(\sqrt{l}+\sqrt{r})^2} \Big(  \dfrac{lr(\sqrt{l}- \sqrt{r})}{(\sqrt{l}+\sqrt{r})}  +\dfrac{1}{2}(l^2- r^2) - \dfrac{1}{4}(\ln l -\ln r) (l + r)^2 \Big)\nu + O(\nu^2),\\
{J}^l_{1}(0) =&   \dfrac{ 3}{2}D_1 \Big(\dfrac{l +  r}{\sqrt{l} + \sqrt{r}} \Big)^2 (l -r) \nu,\\
{J}^l_{2}(0) =&  6D_2 \sqrt{lr}\dfrac{(\sqrt{l}- \sqrt{r})}{\sqrt{l} +  \sqrt{r}} - \dfrac{3(l+r)D_2}{(\sqrt{l}+\sqrt{r})^2} \Big\{ \dfrac{lr(\sqrt{l}- \sqrt{r})}{(\sqrt{l}+\sqrt{r})}\\
&\hspace*{1in} +\dfrac{1}{2}(l^2- r^2) - \dfrac{1}{4}(\ln l -\ln r) (l + r)^2 \Big\}\nu + O(\nu^2).
\end{aligned}
\end{equation}

It can be  verified from equations \eqref{V0I-smallQ} and \eqref{V0I-LargeQ}, that for small and large values of $Q$, the zero-potential current $ I(0)$ is increasing in $Q$ when $l<r$ and it is decreasing in $Q$ if $l>r$. We recall again that the dimensionless quantities can be converted to the dimensional forms from \eqref{1dBV} and Remark \ref{rem-dim2}.

Figure \ref{fig-IQ3} admits the above conclusion. Besides, it suggests that the monotonicity of $ I(0)$ holds for all values of permanent charge, not only for small or large values. We emphasize that the monotonicity of current $I$ with respect to permanent charge $Q$ is just true for zero membrane potential, i.e., $V=0$. Indeed, one should recall from section \ref{sec-IVQ} that when $V \neq 0$, then the current $I$ is not monotonic in $Q$.

\begin{figure}[H]\label{fig-IQ3}
\centerline{\epsfxsize=3.0in
\epsfbox{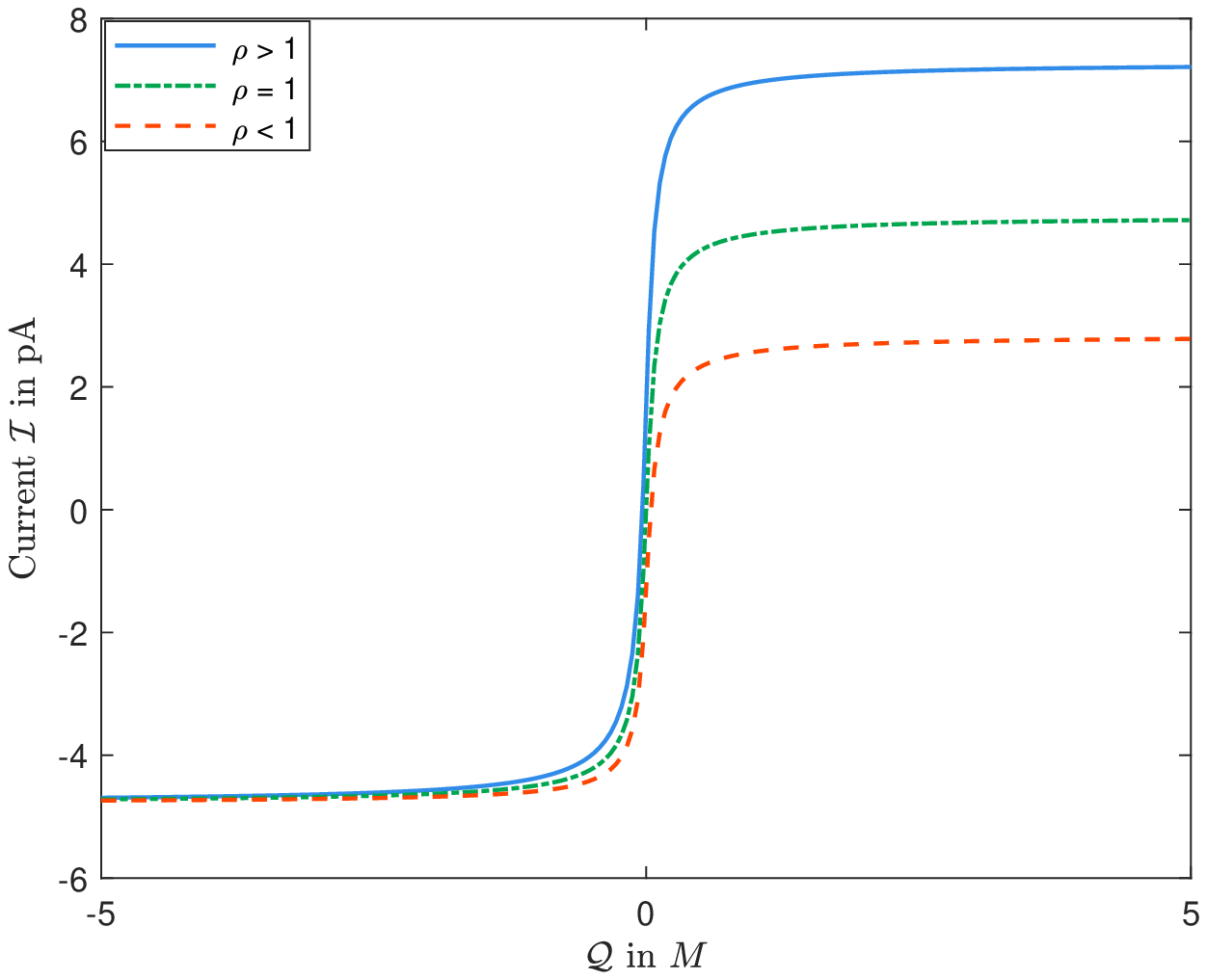} 
\epsfxsize=2.8in
\epsfbox{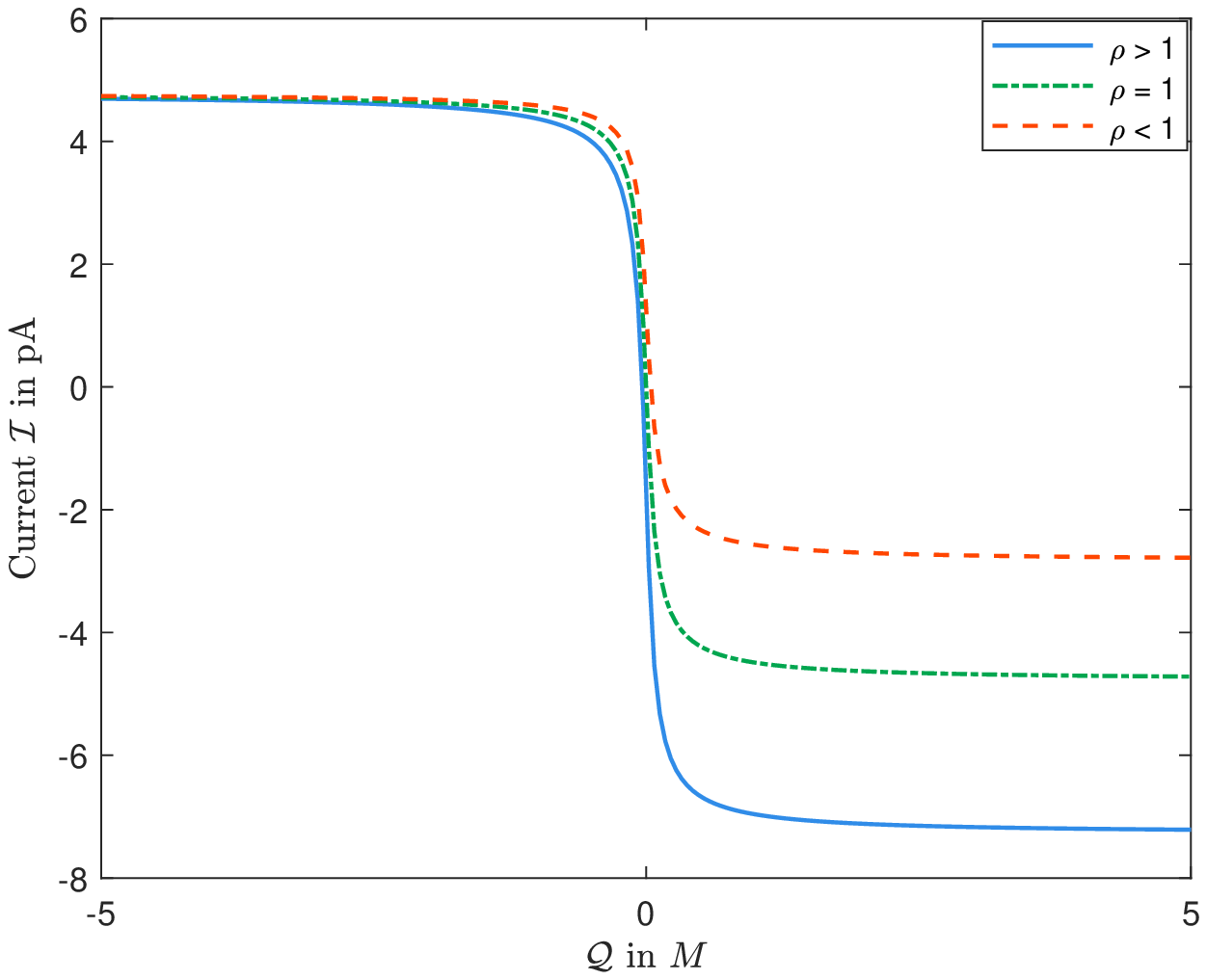} 
  }
\caption{\em The function $\mathcal{I}=\mathcal{I}(\mathcal{Q})$ for $\mathcal{V}=0$: The left panel for $L= 20 mM$ and $R= 30 mM$; the right panel  for $L= 30 mM$ and $R= 20 mM$.}
\label{fig-IQ3}
\end{figure}


\section{Conclusion.}\label{sec-Conc}
 In this paper, we first recall the analytical results in \cite{ML19} for arbitrary diffusion constants.  
 To investigate the reversal potential problems, i.e., when the current is zero, we do numerical investigations based on the analytical results in \cite{ML19}, where many cases are studied analytically. We derive several remarkable properties of biological significance, from the analysis of these governing equations that hardly seem intuitive.

Biophysicists are also interested in the relation of current-voltage (I-V), and current-permanent charge (I-Q), as well as reversal potentials problems. 
To do that, we first recall the analytical results in \cite{EL}, for arbitrary diffusion constants, to drive the flux densities and current equations explicitly. One way to characterize channels is the current at zero potential, that is when $V=0$, that has practical advantages. Since it is usually easier to measure a large current than a vanishing one, we analyzed this case, as well. 
Furthermore, we briefly study the special cases of small and large permanent charge for zero voltage case, based on the analytical results of \cite{JLZ15} and \cite{ZEL} respectively. To bridge between small and large values of permanent charges, we numerically study I-V and I-Q relations  for this case as well.

\section{Acknowledgement.}
The research is partially supported by Simons Foundation Mathematics and Physical Sciences-Collaboration Grants for Mathematicians \#581822.

 \footnotesize
 
  \bibliographystyle{plain}

\end{document}